\documentstyle[12pt, psfig]{article}

\begin{document}
\rm

\begin{center}
\Large{\bf The highest-energy cosmic rays} 
\end{center}
\vspace{0.5in}
\begin{center}
\large{James W. Cronin\\
Center for Cosmological Physics\\
Enrico Fermi Institute\\
University of Chicago\\
5640 S. Ellis Ave\\
Chicago IL 60637\\
USA}
\end{center}
\vspace{0.2in}
\begin{center}
{\em Talk presented at TAUP 2003, Seattle, USA}
\end{center}
\vspace{0.5in}

\noindent{\Large{\bf Abstract}}
\vspace{0.2in}

This is a review of the experimental data concerning the spectrum, arrival
distribution and composition of cosmic rays with energies 
$\geq$ 10$^{19}$ eV. Before the experimental review
I briefly discuss the conditions for the production followed by a review of the
propagation of cosmic rays. Then follows a discussion of the properties of
the showers produced by the primary cosmic ray particles and a description
of the techniques used to detect the showers and extract the energy,
direction and nature of the primary.
The main conclusion of the experimental review is that there is still
insufficient data to answer all the questions 
concerning the particles which strike the
earth with such enormous energies.  There has been significent progress
which I will discuss and there are good prospects that in the
next five years we will come much closer to the answers. Much more can be 
learned from existing data but a more sophisticated and disciplined analysis 
will be required.

\section{Introduction}

Over the past ten years the interest in the nature and origin of the
highest-energy cosmic rays, those with energy $\geq 10^{19}$ eV, has
grown enormously. Of
particular interest are cosmic rays with energy $\geq10^{20}$ eV. At these
energies the cosmic ray particles, be they protons, nuclei, or photons,
interact strongly with the cosmic microwave background and should be
severely attenuated, except for those whose sources are in our cosmological
neighborhood ($\leq$ 100 Mpc). Also protons of these energies may not
be significantly deflected so that some of the cosmic rays may point back
to their source. A recent authoritative review by Nagano and Watson 
\cite{nw00} presents the experimental and theoretical background. For 
all but the most recent references I refer the reader this article.

I dedicate this article to the memory of John Linsley (b. March 12, 1925;
d. September 15, 2002).
He discovered the
first cosmic ray with an energy of $10^{20}$ eV \cite{Linsley1} 42 
years ago. John Linsley
has contributed so much to our understanding of cosmic ray showers. 
Many of his contributions are tucked away in old proceedings of 
cosmic ray conferences. His original development of the concept of
elongation rate is to be found in the Proceedings of the
15th ICRC held in Plovdev, Bulgaria \cite{Linsley2}. (The
elongation rate is referred to in section 3 and section 5.3 of this paper.)
Up until his death he was
active in all aspects of cosmic ray physics. He was the founder of the
idea for a satellite-borne fluorescence telescope, which is being realized
in the EUSO detector to be placed on the Space Station. 

He considered me an upstart when I began
with colleagues to argue for a really large surface detector' one which
ultimately became the Auger Observatory. 
However with the passage of time I believe I gained
his respect. I only wish that he could witness the progress that is 
going to be made. I cannot avoid the fantasy that he is now in a position
to know all the answers!

\begin{figure}
\centerline{\hbox{\psfig{figure=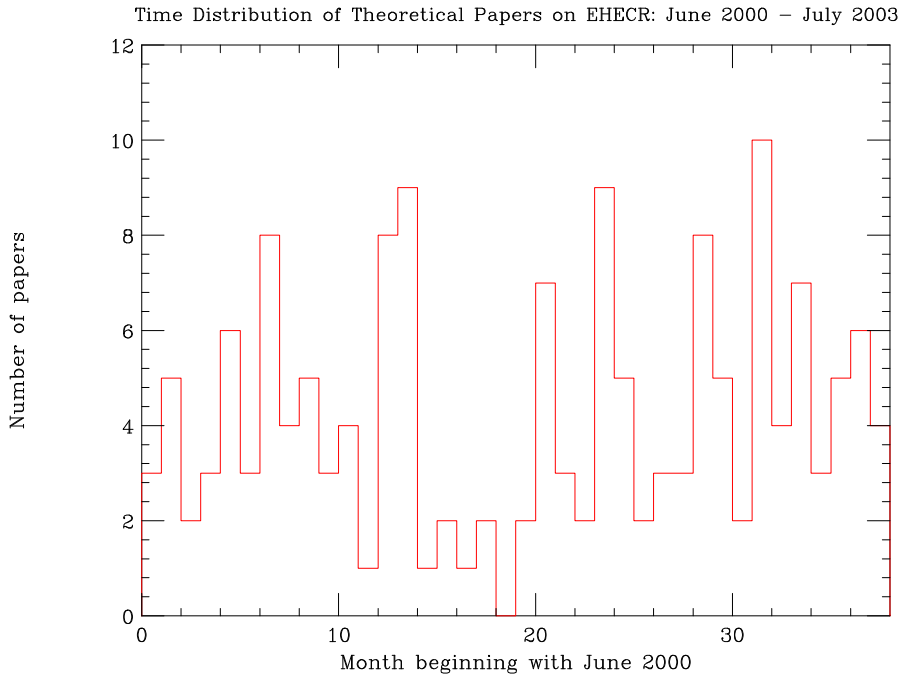,height=3.5in,angle=90}}}
\caption{Number of theoretical and speculative papers on the subject of
the highest-energy cosmic rays.}
\end{figure}

\begin{figure}
\centerline{\hbox{\psfig{figure=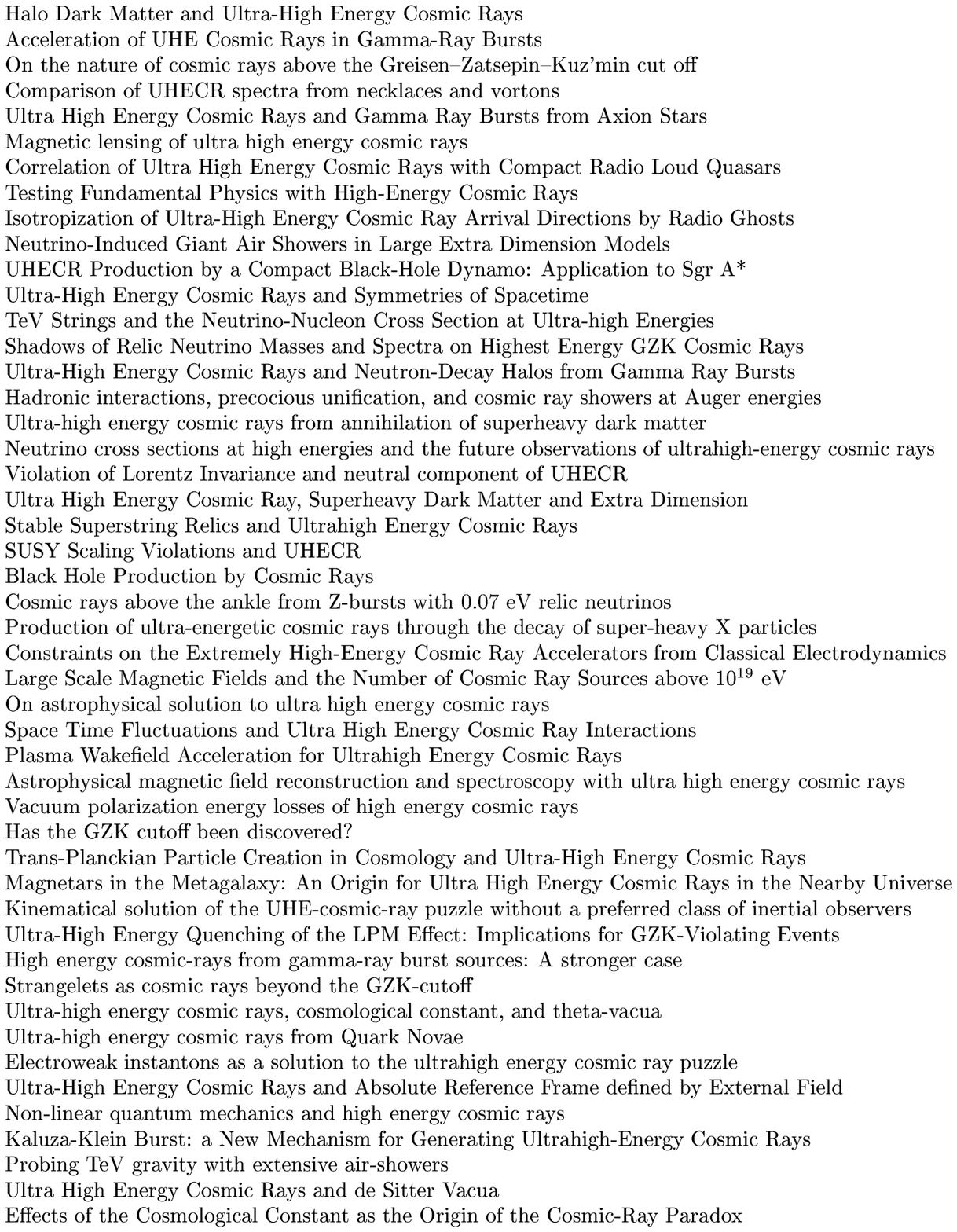,height=8.5in,angle=0}}}
\caption{Some titles of paper concerning the highest-energy cosmic rays.}
\end{figure}

Much more data will be required to understand the mystery behind the
existence of cosmic rays with such extraordinary energies. An article 
by Michael Hillas \cite{Hillas1}, published nearly twenty years ago,
presented the basic requirements for the acceleration of particles
to energies $\geq10^{19}$ eV by astrophysical objects. The requirements
are not easily met, which has stimulated the production of a large
number of creative papers.

In Figure 1 I plot the number of theoretical papers, mostly speculative,
written on the subject of the highest-energy cosmic rays as a function of 
time, as found on the Los Alamos server as astro-ph papers. Over the 
last three years 
the average has been one paper per week. In Figure 2 I list a random sample of
the titles. The authors of these papers deserve a strong response from
the experimental community.

\begin{figure}
\centerline{\hbox{\psfig{figure=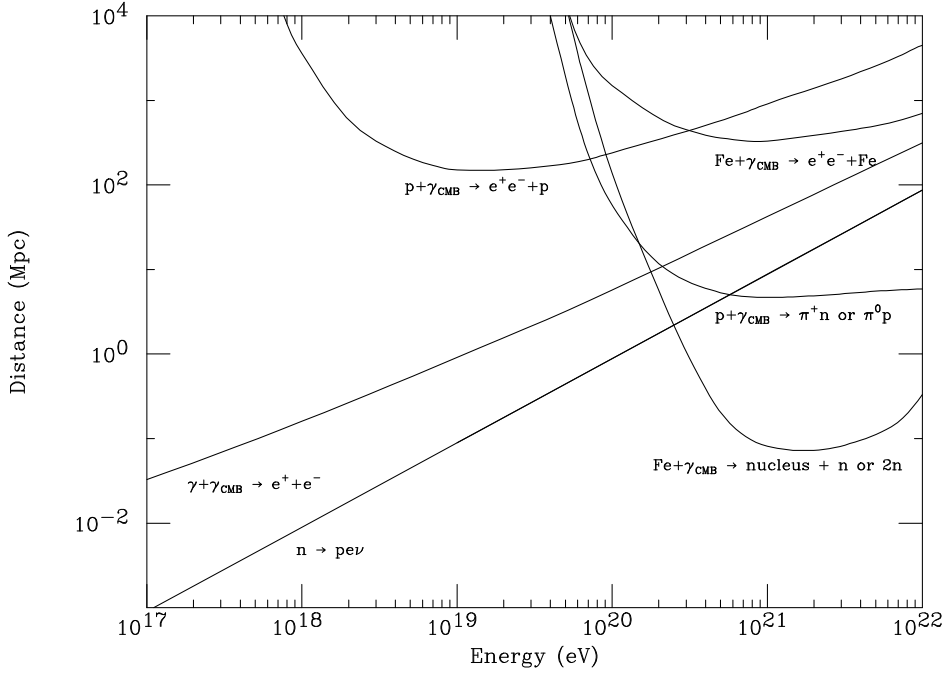,height=3.5in,angle=90}}}
\caption{Panorama of the interactions of possible cosmic primaries with the
CMB. Curves marked by ``p+$\gamma_{CMB}$ $\rightarrow$ e$^{+}$e$^{-}$+p''
and ``Fe+$\gamma_{CMB}$ $\rightarrow$ e$^{+}$e$^{-}$+p'' are energy loss 
lengths
(the distance
for which the proton or Fe nucleus loses 1/e of its energy due to pair
production). The curve marked by ``p+$\gamma_{CMB}$ $\rightarrow$ $\pi^{+}$n
or $\pi^{\circ}$p'' is the mean free path for photo-pion production of a proton
on the CMB. The curve marked ``Fe+$\gamma_{CMB}$ $\rightarrow$ 
nucleus + n or 2n''
is the mean free path for a photo-nuclear reaction where one or two nucleons are
chipped off the nucleus. The curve marked 
``$\gamma$ +$\gamma_{CMB}$ $\rightarrow$ e$^{+}$e$^{-}$'' is the mean free 
path for the interaction of a high-energy photon
with the CMB. Added for reference is the mean decay length for a neutron 
indicated by ``n $\rightarrow$ pe$\nu$''.}
\end{figure}

\section{Propagation of the highest-energy cosmic rays}

The interaction of the particles with the
cosmic microwave background (CMB) and magnetic fields plays an important role
in their propagation. All possible species of cosmic rays with the exception
of neutrinos interact with the CMB. A panorama of the various interactions 
is given in Figure 3.

\begin{figure}
\centerline{\hbox{\psfig{figure=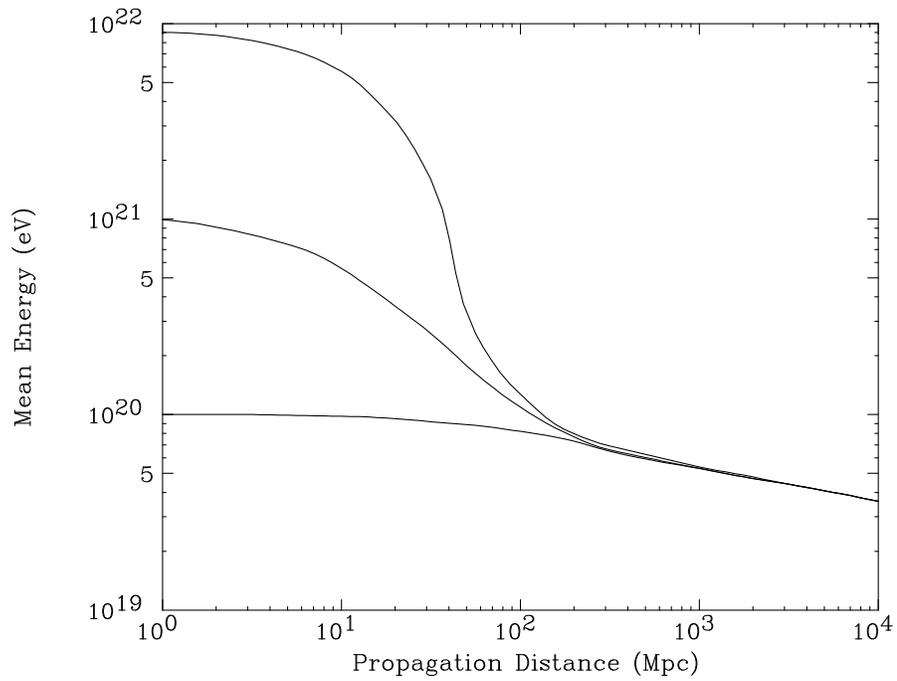,height=3.5in,angle=90}}}
\caption{Mean energy of protons as a function of propagation distance through
the CMB. Curves are for energy at the source of 10$^{22}$ eV, 10$^{21}$ eV,
and 10$^{20}$ eV.}
\end{figure}

\begin{figure}
\centerline{\hbox{\psfig{figure=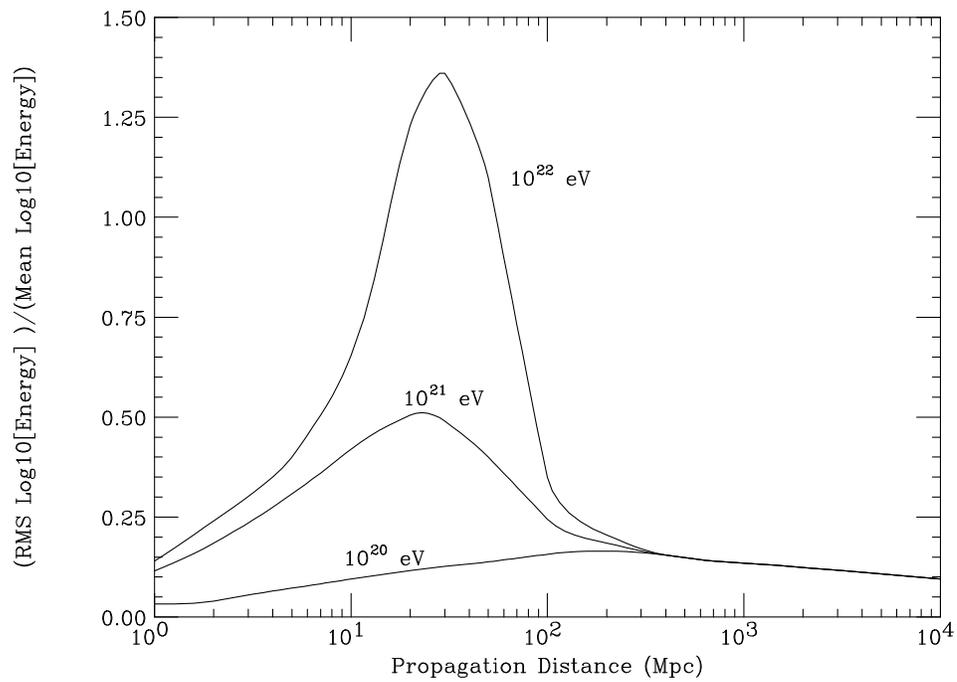,height=3.5in,angle=90}}}
\caption{Fluctuation of the energy of a proton propagating through the
CMB.}
\end{figure}

The most discussed is the Greisen-Zatsepin-Kuzmin (GZK) 
effect \cite{Greisen}, where a cosmic ray proton interacts with a CMB
photon. The collision of a $10^{20}$ eV proton with a $10^{-3}$ eV photon
produces about 200 MeV in the center of mass, which is the peak for
photo-pion production. These collisions produce a neutron and charged 
pion or a proton and neutral pion with significant loss of energy for the
nucleon. The neutron mean decay length is 1 Mpc at 10$^{20}$ eV so on
the Mpc scale it quickly becomes a proton again. The mean energy of a
nucleon as a function of propagation distance through the CMB is shown in
Figure 4. The interaction of a nucleon with the CMB is a stochastic one 
where a significant amount of energy can be lost in a single collision.
As a consequence the fluctuation of the energy of a nucleon about the mean
as a function of distance is large. The fluctuation about the mean is
shown in Figure 5.

\begin{figure}
\centerline{\hbox{\psfig{figure=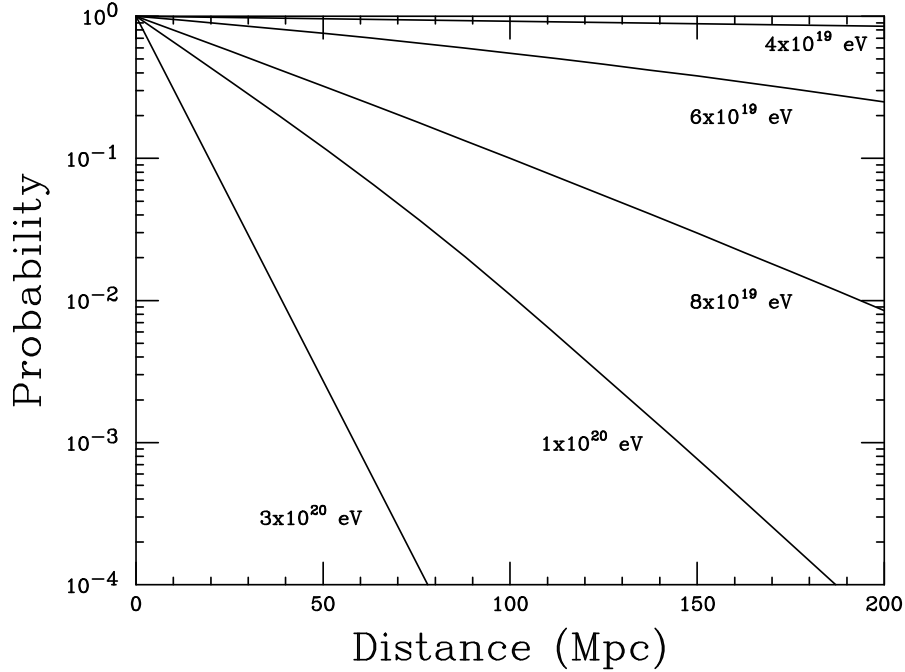,height=3.5in,angle=90}}}
\caption{Probability that an observed event at a given energy has its
source at a distance greater than the indicated distance. A source spectrum
proportional to E$^{-2.5}$ is assumed.
Figure provided Paul Sommers, University of Utah.}
\end{figure}

While these two figures present the physics of propagation of protons
in the CMB, the observer has to answer the inverse question. If a
cosmic ray is observed with a particular energy, what is the probability
that it came from a distance greater than a specified amount? To compute
this probability requires an assumption about the spectrum at the source.
With  an assumed source spectrum of E$^{-2.5}$ the probability is
plotted in Figure 6. The GZK interaction begins to have a significant
effect at an energy of 8 x 10$^{19}$ eV where there is  only a 10$\%$
probability that the cosmic ray traveled a distance greater than
100 Mpc. Cosmic rays at lower energies can traverse much greater distances
with little energy loss, but it is also possible that such a cosmic ray
was produced at a higher energy and encountered a CMB photon at a great 
distance. 

Since the energy loss in the collisions with the CMB has such large
fluctuations, a set of data with a small number of events above 10$^{20}$
eV will have fluctuations far in excess of simple Poisson statistics. The
importance of these fluctuations has been discussed in a recent paper 
\cite{Olinto},
which demonstrates that none of the experiments done up to the present
have the statistical power to address the existence of a GZK cutoff. But
I must stress that a single event observed with energy 3 x 10$^{20}$ eV has
about a 0.1$\%$ chance of having traveled more than 50 Mpc.  

The expectation of the number and spectrum of post-GZK events is not a simple 
calculation, since the distribution of possible sources within 100 Mpc may 
not be treatable as a continuum. The distribution
of nearby sources is certainly not identical as one looks in different 
directions of the sky. For this reason, a comprehensive study of these 
highest-energy cosmic rays requires the
observation of the entire sky. One cannot assume that the spectrum at its
upper end is independent of the direction of observation.

A major portion of the cosmic rays is charged particles. There is evidence,
not yet conclusive, that protons are the predominant species above 
10$^{19}$ eV.  
The Larmor radius of a charged particle with atomic number Z is:

\begin{center}
{$\rho$(Mpc)=1.08x10$^{2}$$\cdot$(E/10$^{20}$eV)$\cdot$(1/B$_{ng}$)$\cdot$(1/Z).} 
\end{center}

A proton of 10$^{19}$ eV in the galactic magnetic field of about 1 $\mu$-gauss
has a Larmor radius of 10 kpc, large compared to the thickness of the galaxy.
Thus, if a proton of such an energy is born in the galactic disk, it will 
immediately escape. Iron nuclei of that energy will linger in the galaxy
but will show an anisotropy if they are born in the galactic disk.
As there is no correlation of the arrival directions of $\geq$ 10$^{19}$
eV  cosmic
rays with the galactic disk it appears that their sources are either 
extragalactic or originate in an extended halo of the galaxy.

\begin{figure}
\centerline{\hbox{\psfig{figure=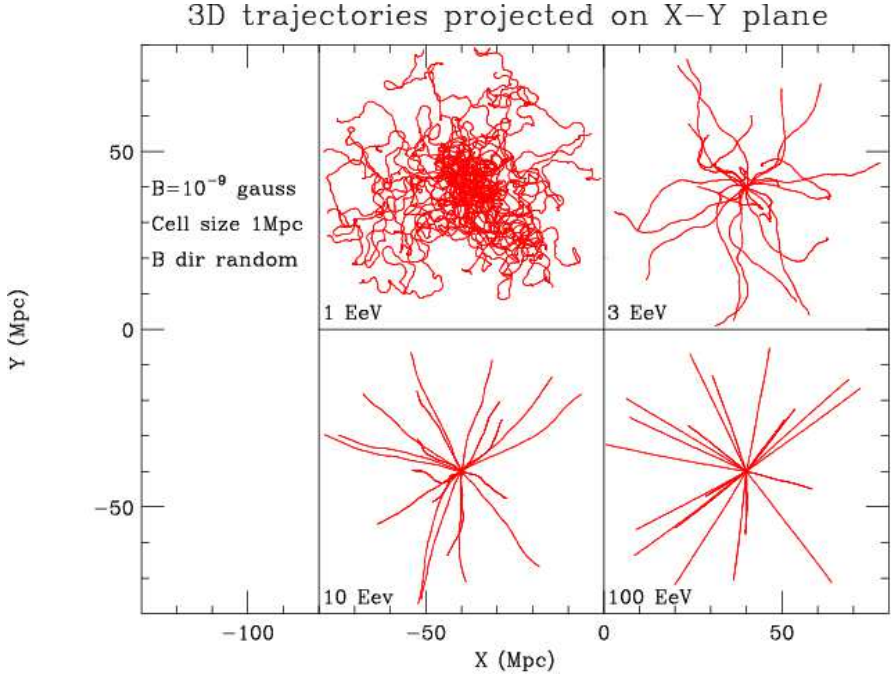,height=3.5in,angle=90}}}
\caption{Projected view of 20 trajectories of proton primaries emanating
from a point source for several energies. Trajectories are plotted until
they reach a physical distance from the source of 40Mpc. See text for details.}
\end{figure}

Not a great deal is known about magnetic fields outside of galaxies. In
clusters of galaxies such as the Virgo cluster the fields can be of the
order of a $\mu$-gauss \cite{Kronberg1}.
With exaggerated
symplicity these fields are often described by  cells of size $\lambda$ 
= 1Mpc with mean value B oriented randomly in each cell. This is the model 
taken for the measurement of the magnetic fields by Faraday rotation 
\cite{Kronberg2}. 
In the voids or near voids of extragalactic space an estimate of the
magnitude of the randomly oriented magnetic field is $\leq$ 1 nanogauss.  

I have made some simple calculations to give a feeling of the very important
effects that even such a small field can have on the propagation of charged
cosmic rays \cite{Cronin}. In Figure 7 I have plotted in plane projection
twenty 3-dimensional trajectories for cosmic ray protons emitted 
from a point source. The
trajectories are followed until they reach a spatial distance of 40 Mpc from
the source. The actual traversed distance which is relevant for the various
energy loss processes can be much longer. No energy loss was applied in these
calculations. One can see that for the assumed magnetic conditions the
propagation of the cosmic rays passes from diffusive propagation to 
rectilinear propagation in passing from 1 EeV to 100 EeV. (I hope the reader
will excuse the shift in energy units here, 1 EeV = 10$^{18}$ eV) These graphs
can be easily scaled for other magnetic conditions. For example, if the
magnetic field were 100 nanogauss, propagation at 100 EeV would be completely
diffusive, as shown in the upper left panel of Figure 7.  Propagation at 1000 
EeV however would be quite distinct from the lower left panel as energy loss
by the GZK effect would be significant. Less than 1$\%$ of the particles
would escape interaction with the CMB and propagate rectilinearly. The remainder
would quickly pass to diffusive propagation, drop below 100 EeV, and travel
much more slowly from the source. 
For iron primaries, the panel on the upper right of Figure 7 would correspond 
to 80 EeV. This regime is not fully diffusive and the primaries would have some
memory of their source which would be revealed by a broad anisotropy.  
These examples reveal the complexity introduced in propagation of cosmic
rays due to magnetic fields. In some cases the galactic magnetic field will
also be important.

\begin{figure}
\centerline{\hbox{\psfig{figure=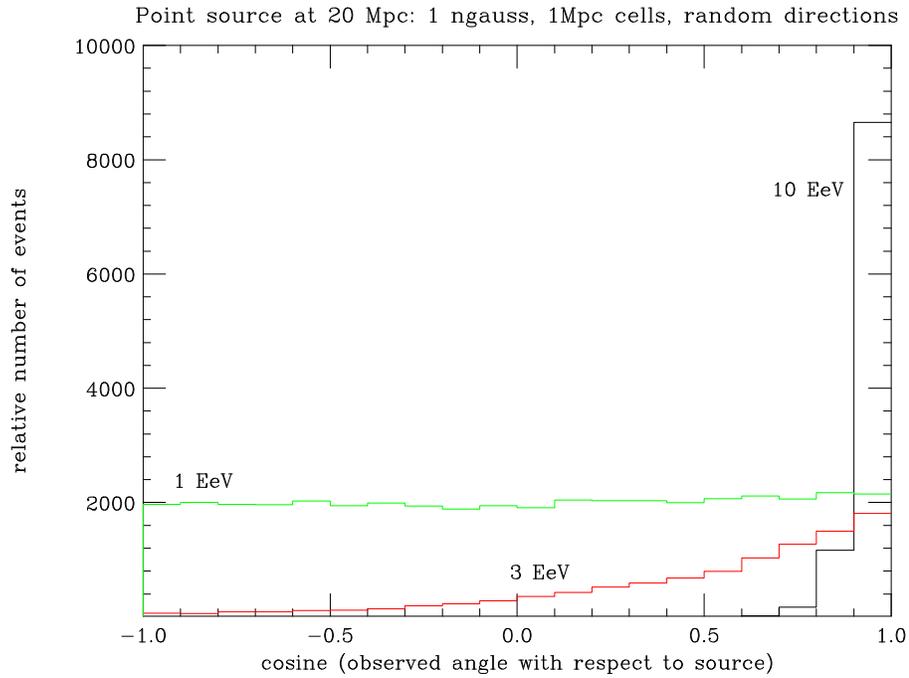,height=3.5in,angle=90}}}
\caption{Distribution at 20 Mpc distance of the observed cosmic ray angle 
with respect to the source direction for proton primaries at the indicated 
energies.}
\end{figure}

\begin{figure}
\centerline{\hbox{\psfig{figure=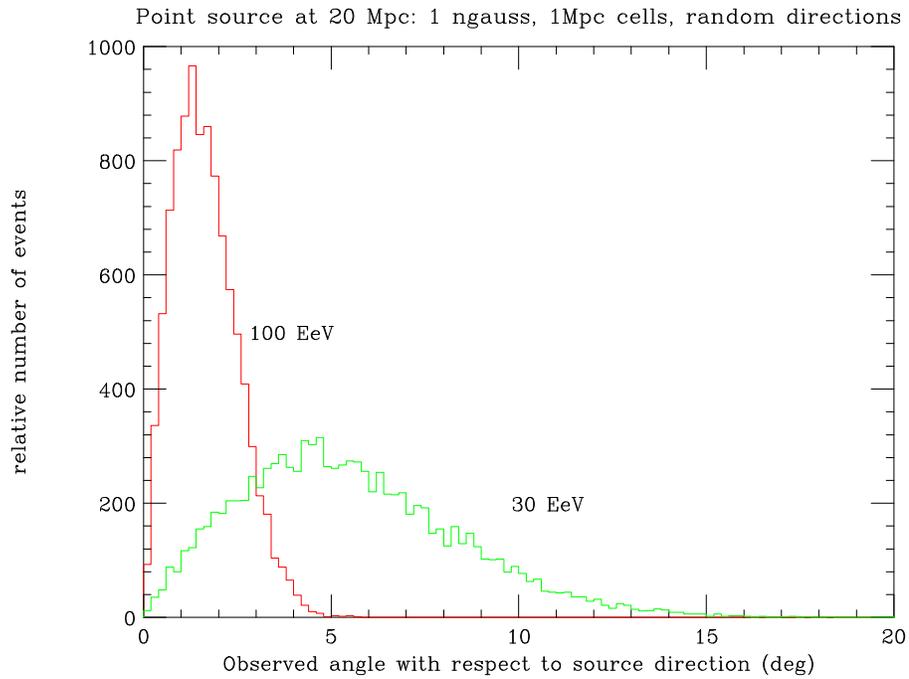,height=3.5in,angle=90}}}
\caption{Distribution at 20 Mpc distance
of the observed cosmic ray angle with respect to
the source direction for proton primaries for the cases of small deflection.}
\end{figure}

\begin{figure}
\centerline{\hbox{\psfig{figure=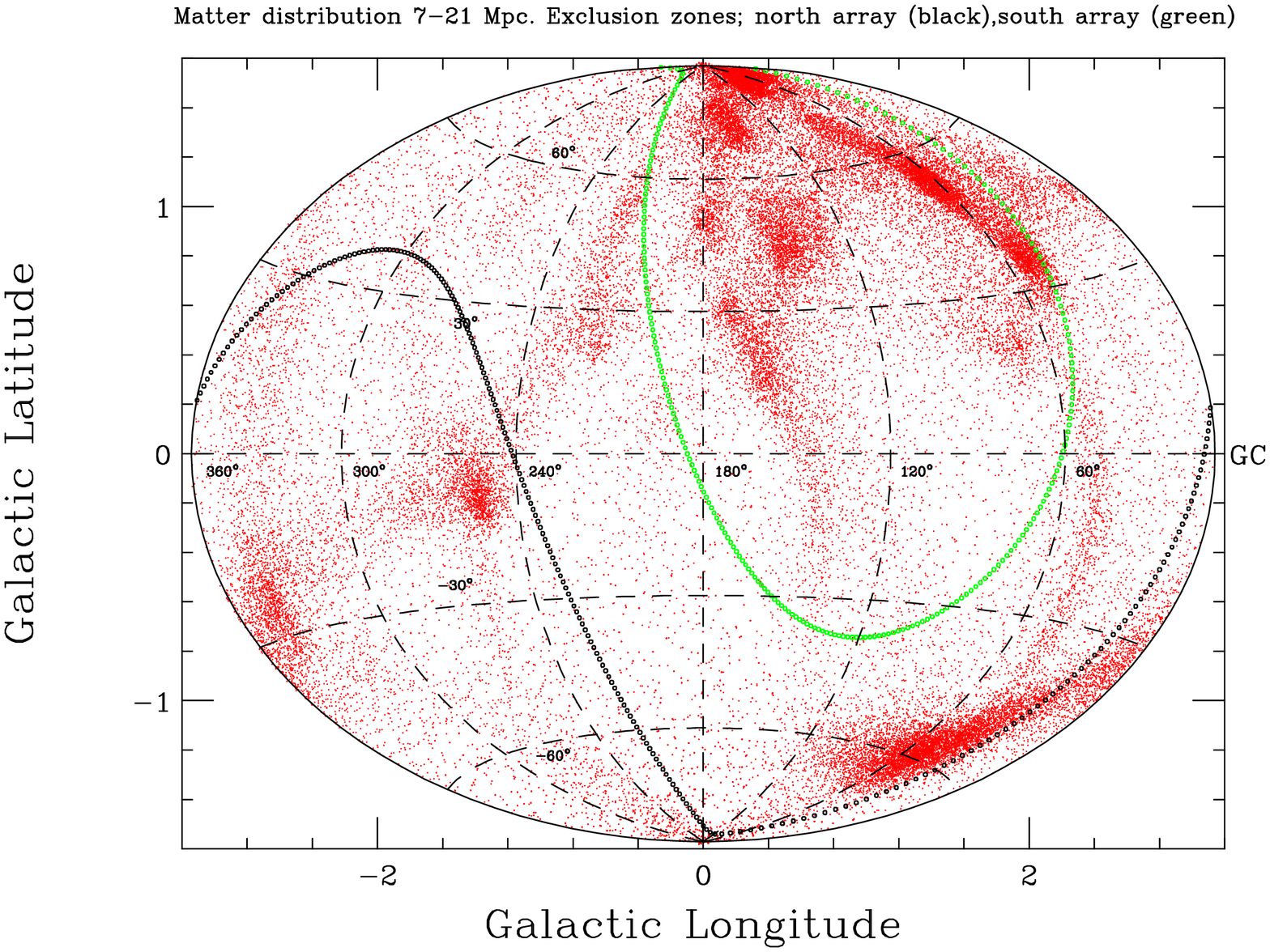,height=4.0in,angle=90}}}
\caption{Distribution in galactic coordinates of gravitating matter at
distance between 7 and 21 Mpc. The density of dots in the figure is proportional
to the column density of the matter in the interval. The exclusion zones are 
plotted for a zenith angle $\leq$ 60$\deg$. On the left is the exclusion zone
for a northern observatory, on the right for a southern observatory. 
An observatory in the southern 
hemisphere will not see directly the Virgo cluster. (The data were provided
by Andrey Kravtsov, Center for Cosmological Physics, University of Chicago).}
\end{figure}

In Figure 8 I have plotted the distribution of observed directions of the
cosmic rays with respect to the source direction. For 1 EeV proton primaries
the directions are completely isotropic; no memory of the source direction
remains. In Figure 9 I plot the dispersion of angles for 100 EeV and 30 EeV
proton primaries. Here the angular spread is 1.5$^{\circ}$ and 5$^{\circ}$ 
respectively.

If the sources of cosmic rays with energy $\geq$10 EeV are extragalactic and
are associated with the distribution of nearby matter, then one would expect
that the flux and energy spectrum of the cosmic rays will depend on the
hemisphere in which the observations are made. Most of the nearby matter
is found in the Virgo cluster at a distance of $\sim$ 18 Mpc. In Figure 10
I plot the column density of gravitating matter between 7 and 21 Mpc. Also 
plotted in the
figure are the exclusion zones for observatories at 35$^{\circ}$ north and south
latitude. The bulk of the Virgo cluster is not seen by an observatory in the
southern hemisphere.

\section{Properties of showers}

When a cosmic ray strikes the earth's atmosphere a shower of particles
is produced. In this section I describe the properties of these showers
with emphasis on the features that are important for the detection, and
measurement of the energy, direction, and nature (proton, nucleus, or photon)
of the primary cosmic ray. These properties are derived from both simulations
\cite{Simulations}
and data from the Engineering Array of the Pierre Auger Observatory
\cite{Auger}. This discussion concerns showers with zenith angles of less
than 60$^{\circ}$. Steeper showers, often referred to as horizontal
showers, begin to be dominated by muons as the electromagnetic component
quickly dies out. The method to analyse horizontal showers is discussed
in an Auger Observatory paper \cite {Ave}, and is beyond the scope of 
this paper.

\begin{figure}
\centerline{\hbox{\psfig{figure=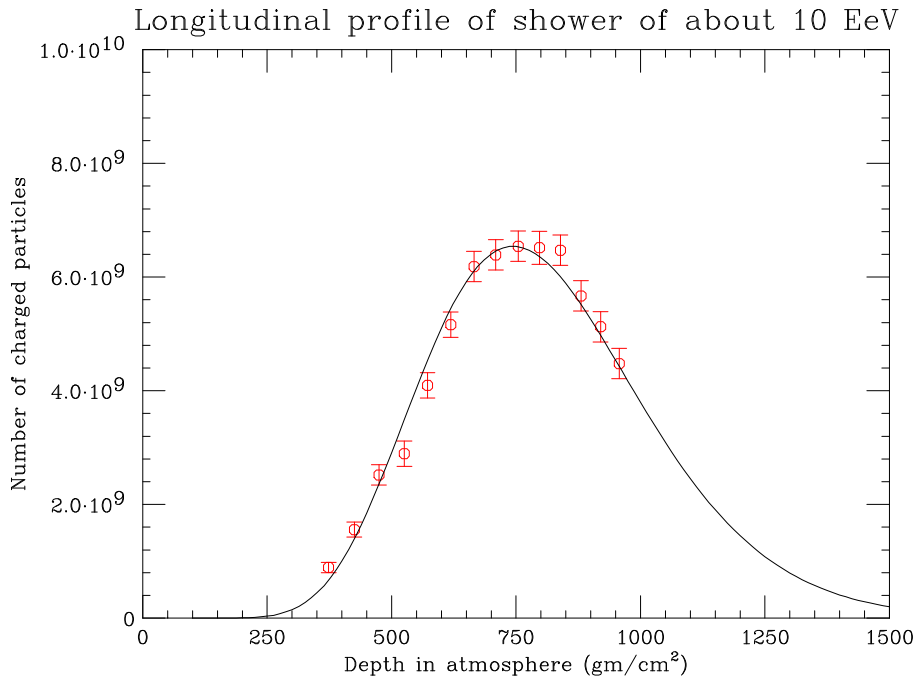,height=3.5in,angle=90}}}
\caption{Longitudinal development of a 9 EeV shower in the atmosphere.
The solid line is an empirical fit to the shower development. The red
points are the measurements of the shower development by detection of
its nitrogen fluorescence.}
\end{figure}

In Figure 11 we show the longitudinal development of a shower of energy
about 10 EeV. The solid curve is an empirical shape proposed by Gaisser
and Hillas \cite{Gaisser-Hillas}. The points are a measurement by one 
of the fluorescence telescopes of the Pierre Auger Observatory. As will be
discussed below the longitudinal development in the atmosphere can be
directly measured by observation of nitrogen fluorescence. 
A 10$^{19}$ eV shower produces about 7 x 10$^{9}$ charged particles at its 
maximum 
development. Most of the shower particles are very close to an axis defined
as the direction of the primary particle. Roughly
80$\%$ of the particles are within one Moli\`{e}re radius of the shower axis.
This radius, measured in radiation lengths, is the ratio of the multiple 
scattering constant, 21.2 MeV to the critical energy in air, 80 Mev. Typically
this distance is about 100 meters.

\begin{figure}
\centerline{\hbox{\psfig{figure=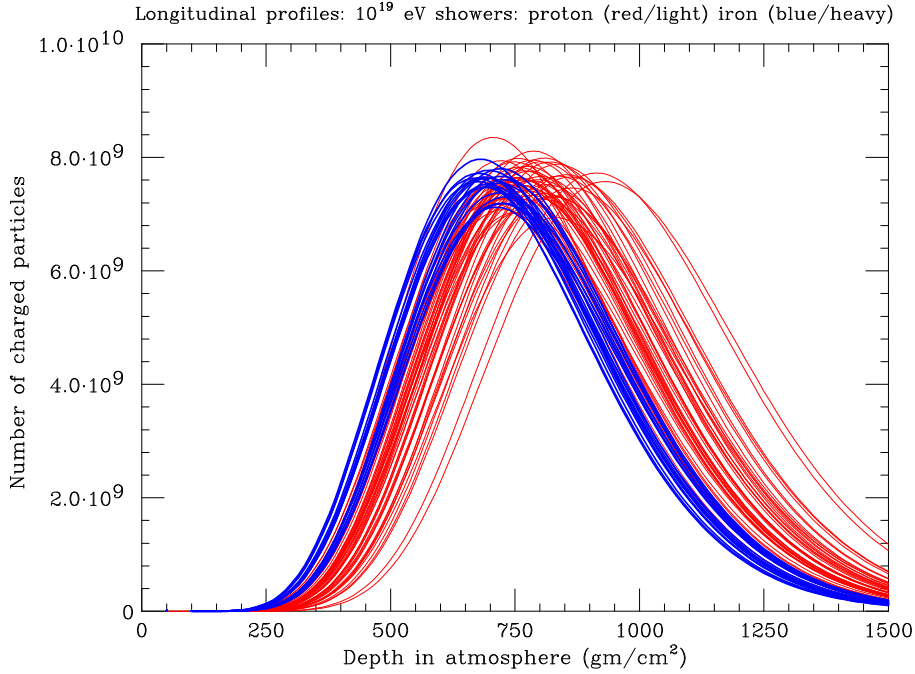,height=3.5in,angle=90}}}
\caption{Simulated shower profiles for proton (light profiles) and iron 
(heavy profiles).}
\end{figure}

The energy of the shower can be obtained by integration of the ionization
loss in the atmosphere, given, approximately, by 2.2$\int$$N_{e}$(x)dx MeV, 
where 2.2 MeV/gm/cm$^{2}$ is dE/dx for electrons at the critical energy, 
$N_{e}$(x) is the number of shower particles at the atmospheric depth x 
gm/cm$^{2}$. 
Implicit in this formula is the presumed knowledge of the shape of the
longitudinal development outside the range of measurement. Also some energy is 
not observed in the form of neutrinos and penetrating muons. Accounting for
this energy adds about 7$\%$ to the ionization energy for a 10$^{19}$ eV shower.

The position of the shower maximum, X$_{max}$, fluctuates principally because
the depth of the first interaction fluctuates. For a proton induced shower
the fluctuation is the largest. For a heavy nucleus like iron, X$_{max}$
is less and the fluctuations are less. This is best understood by considering
an iron shower being produced by 56 nucleons of 1/56th the energy. As discussed
in the following paragraph, X$_{max}$ grows with energy. Thus, 
for an iron shower,
X$_{max}$ is less and the fluctuations are less because one has the
averages over 56 showers. Simulations show that conclusions derived from
this superposition model are remarkably accurate. The model permits
one to make simple semi-quantitative arguments for the comparison of shower
properties. Reading the cosmic ray literature might suggest that the only
hadronic components of cosmic rays are protons and iron nuclei. Of course
this is not the case as many nuclear species between proton and iron are
found in the cosmic rays. Protons and iron represent the extremes. A
number of simulated proton and iron shower profiles is shown in Figure 12. The 
X$_{max}$ for the iron showers is less deep and the fluctuations for the iron 
showers are smaller. The mean X$_{max}$ and its standard deviation for 
protons is 780 and 53 gm/cm$^2$, for iron nuclei 700 and 22 gm/cm$^2$.

\begin{figure}
\centerline{\hbox{\psfig{figure=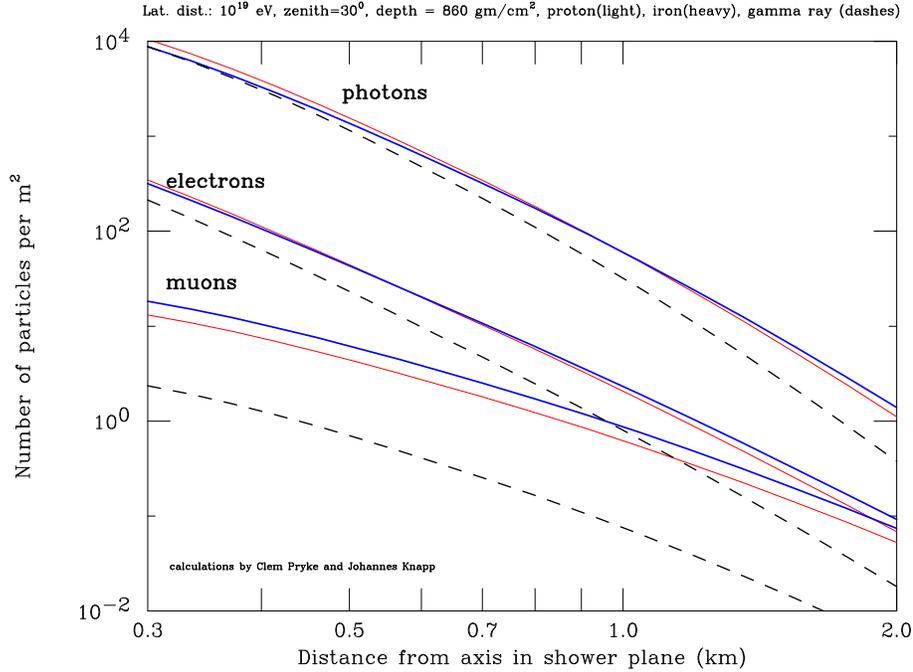,height=3.5in,angle=90}}}
\caption{Simulations of the lateral distribution of
shower particles. The curves are averages over many showers. These are the
densities in the plane perpendicular to the axis of  10$^{19}$ eV proton,
iron, or gamma ray initiated showers. }
\end{figure}

An important quantity obtained from the longitudinal shower development is
the elongation rate dX$_{max}$/dlogE, the increase in X$_{max}$ per
decade. This concept was first introduced by John Linsley \cite{Linsley2}. 
For pure electromagnetic showers, X$_{max}$ = X$_{0}$ln(E/E$_{c}$)
so that  dX$_{max}$/dlogE = 2.3X$_{0}$ = 84 gm/cm$^2$/decade. Here 
E$_{c}$ = 80 MeV is the critical energy in air and X$_{0}$ = 36 gm/cm$^2$
is the radiation length in air.
For hadron primaries with a multiplicity that grows with energy, the elongation
rate is much less. To illustrate the point consider a model in which the
first interaction produces N(E) neutral pions. This is like 2N electromagnetic
showers of energy E/2N. Then X$_{max}$ = X$_0$ln(E/2NE$_c$) and
dX$_{max}$/dlogE = 2.3X$_0$(1-dlnN/dlogE). The rate of increase in the
depth of X$_{max}$ is decreased. Elongation rates around 10$^{19}$ eV are about
60 gm/cm$^2$/decade. The superposition model would predict difference in
X$_{max}$ between iron and proton to be log(56) x 60 = 100 gm/cm$^2$ 
compared to 80 in the simulations. Both the mean value of X$_{max}$ and its 
fluctuation potentially carry information about the primary composition.

\begin{figure}
\centerline{\hbox{\psfig{figure=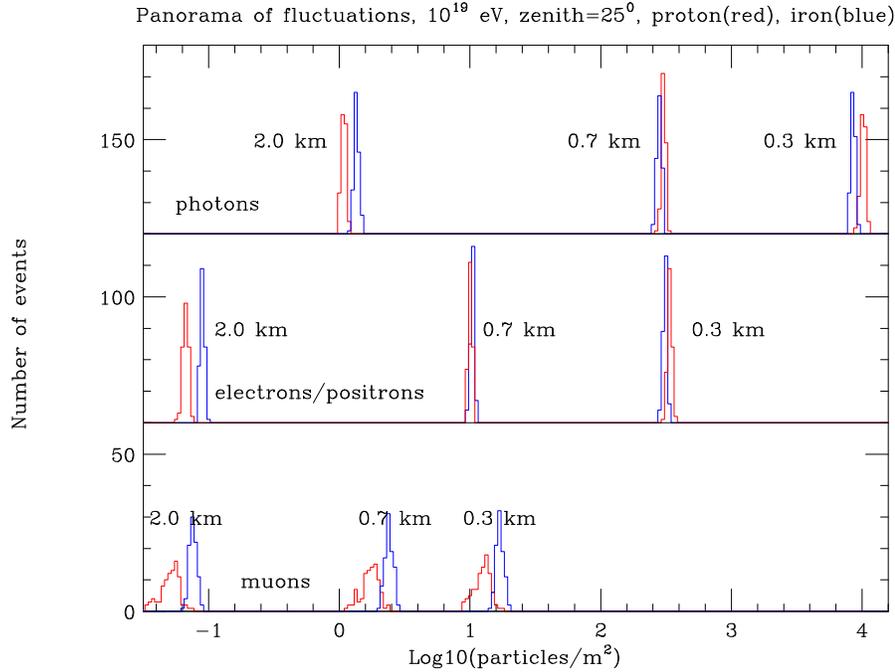,height=3.5in,angle=90}}}
\caption{Density fluctuations of shower particles for selected distances
from the shower axis.  The light histograms 
are for protons and the heavy histograms iron.} 
\end{figure}

An array of particle detectors spread out on the ground is sensitive to the
lateral distribution of the shower particles. The lateral distributions
for showers initiated by protons and iron nuclei are shown in Figure 13.
These curves are the result of simulations. While the quantitative results
depend somewhat on models the qualitative result is unchanged. 
Plotted are the average distributions of 100 showers. The principal components
are photons, electrons and muons. There are hadrons near the shower core
but their contribution is negligible at more than a few 100 meters. One
should note that the density of photons and electrons is almost identical for
proton and iron initiated showers in the range from 500 to 1000 meters from 
the core. The iron initiated showers develop higher in the
atmosphere so the particles are spread more broadly than in a proton 
initiated shower. However 
as the iron initiated shower particles have a greater distance to travel 
than the proton initiated ones there is more attenuation. The two effects 
compensate one another. The gamma ray initiated showers are distinctly
different. These showers have a much steeper lateral distribution because
they develop much deeper in the atmosphere and they have many fewer muons.   
These characteristics make the identification of a gamma ray initiated 
shower easier.
The gamma ray showers have muon densities about a factor five lower than
protons of the same energy. The exact factor depends on the strength of
the gamma ray production of hadrons. This strength has to be extrapolated from
lower-energy data. Nevertheless, one expects many fewer muons in gamma ray
initiated showers.

The muon densities are different for protons and iron, but not by a large
amount.
Iron initiated showers produce
about 1.45 times more muons than proton initiated showers \cite{agasa1}. 
This ratio can be understood qualitatively through
the superposition model. Empirically it is known that the ratio
of muons to electromagnetic particles increases more slowly than the energy.
Typically this ratio goes as E$^{.92}$. Taking iron showers to be 56 proton
showers each of 1/56th the energy, one finds the ratio to be $\sim$ 
56$^{1-.92}$ = 1.38, a reasonable agreement. Simulations
using different interaction models predict electromagnetic densities that
vary at most by perhaps 20$\%$. However different models can predict muon 
densities which can vary by as much as a factor 2.  The qualitative conclusion
however remains that the muon to electromagnetic ratio increases with 
atomic number.

At large distances from the core the intrinsic density
variations  due to fluctuation in the shower maximum are rather small.
Examples of these fluctuations are plotted in Figure 14. The fluctuation
of the electromagnetic component is about 4$\%$ RMS. The fluctuation of 
the muon component is larger being about 20$\%$ RMS for proton induced showers
and 10$\%$ RMS for iron induced showers. These intrinsic fluctuations are 
generally not the limiting factor which determines the precision of the 
air shower detectors.

\begin{figure}
\centerline{\hbox{\psfig{figure=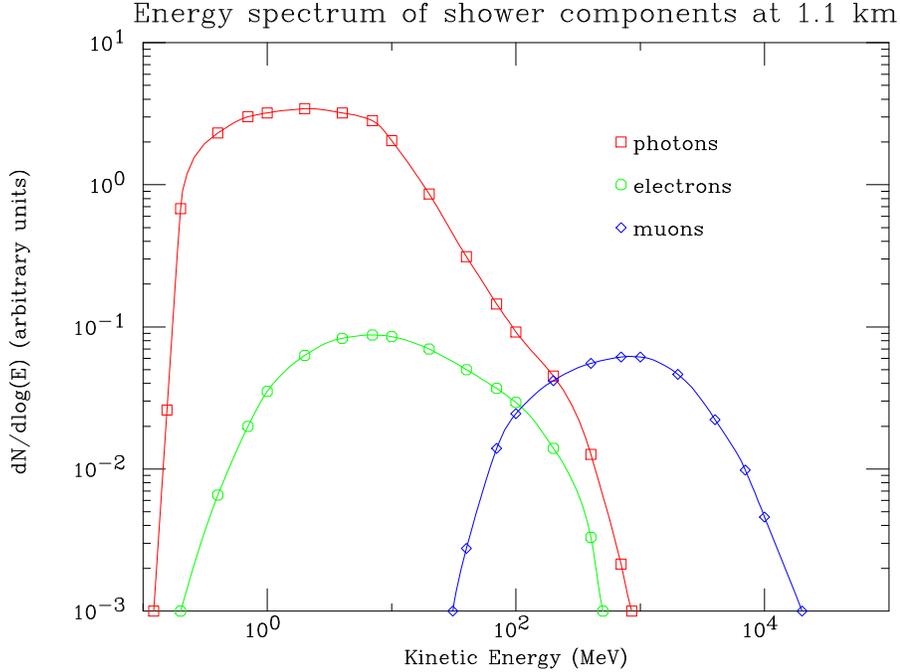,height=3.5in,angle=90}}}
\caption{Energy spectrum of the shower components at 1.1 km. The spectrum
of the electromagnetic components does not vary much beyond 1 km. At 2 km
the peak of the muon spectrum is shifted from 1 GeV to 600 MeV.}
\end{figure}

\begin{figure}
\centerline{\hbox{\psfig{figure=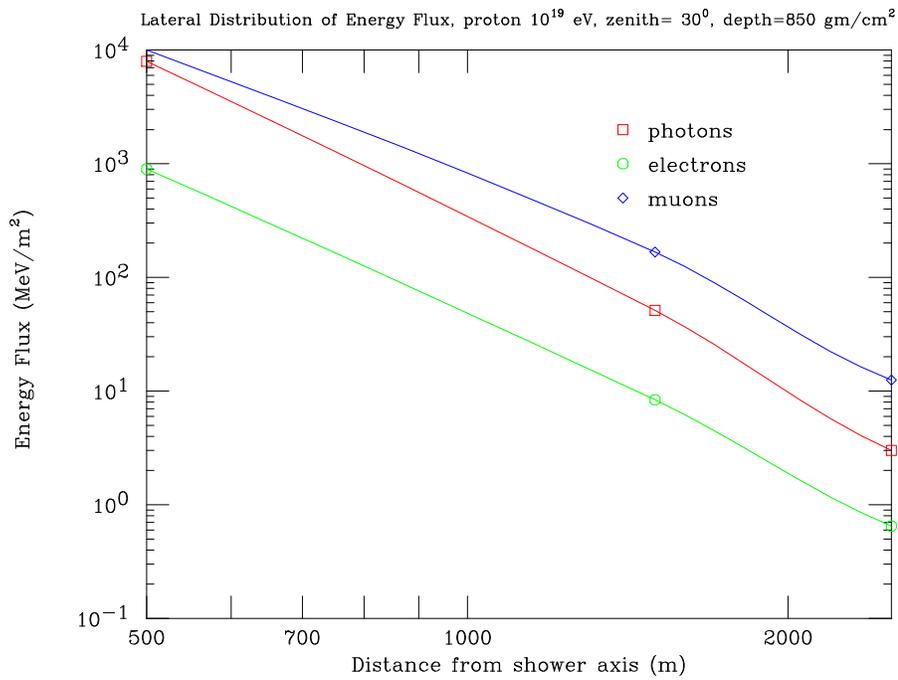,height=3.5in,angle=90}}}
\caption{The lateral distribution of the integrated energy flux for each shower
component.}
\end{figure}

\begin{figure}
\centerline{\hbox{\psfig{figure=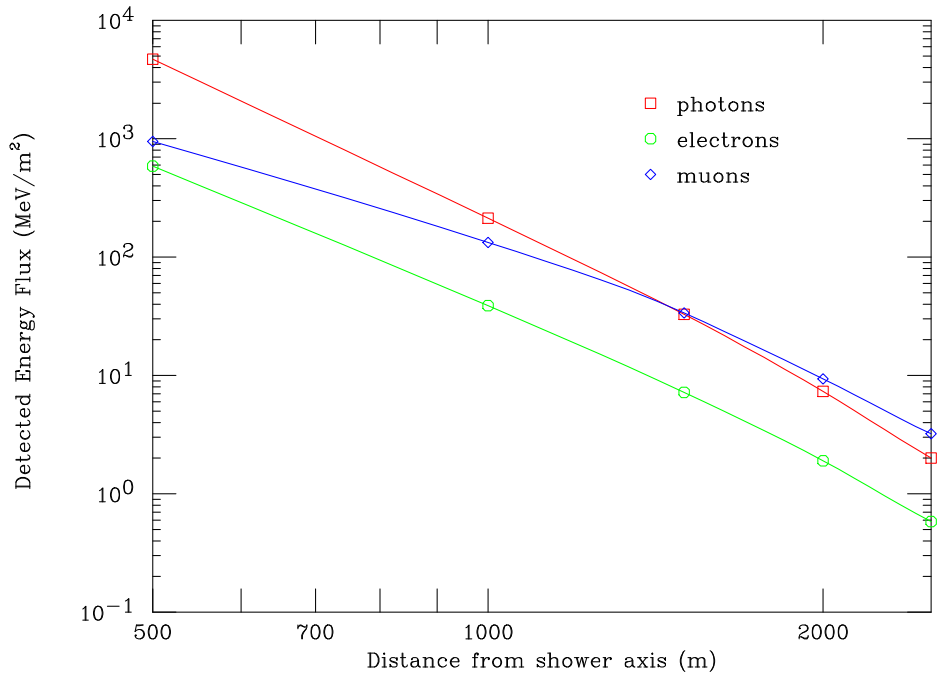,height=3.5in,angle=90}}}
\caption{The lateral distribution of the detected energy flux for each shower
component.}
\end{figure}

A scintillator detector responds to the number of ionizing particles that 
pass through it. If there is little material above, the scintillator records 
only the electrons and muons. The density of particles at 600 meters is
related to the primary energy and, following Figures 13 and 14,
this density is relatively insensitive to the nuclear composition.

Tanks of water have also been used as particle detectors using the
Cherenkov radiation produced by the charged particles \cite{watson}.
They are 
also very efficient in detecting the photon component of the shower.
A deep water detector responds more directly to the energy flux of the shower 
particles. The energy distribution of the shower particles is shown in 
Figure 15. The integrated energy flux as a function of distance from the shower
axis is shown in Figure 16. The energy flux far from the core is carried
principally by the muons. The energy flux actually detected in 1.2 meters
of water is shown in Figure 17. Water tanks of 10 m$^{2}$ in area and 1.2 m
deep will be used as surface detectors in the Auger Observatory. This shape
has the property that the average energy deposit of each shower component is
nearly independent of the zenith angle out to 60$^{\circ}$.

\begin{figure}
\centerline{\hbox{\psfig{figure=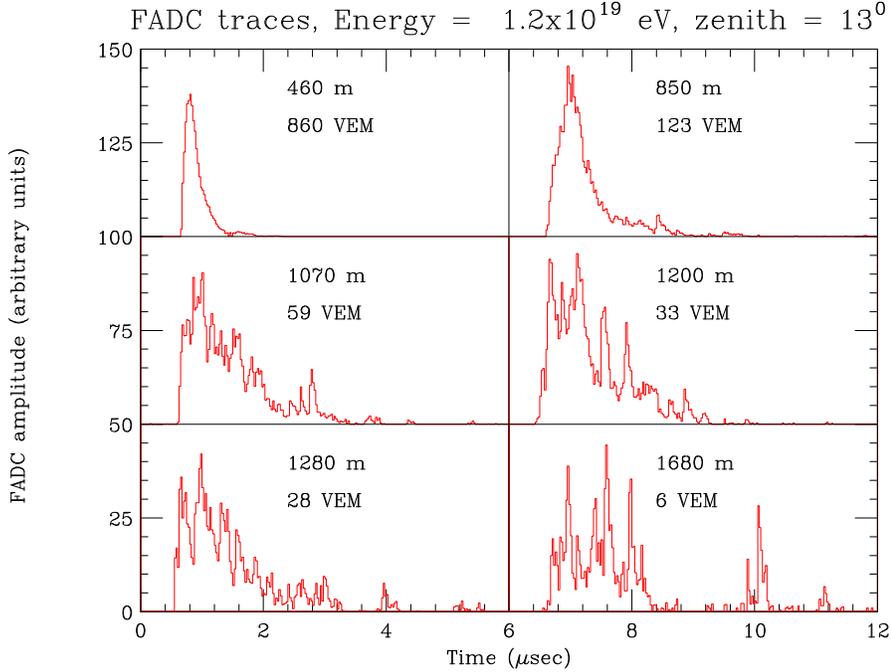,height=3.5in,angle=90}}}
\caption{FADC traces for a $\sim$10$^{19}$ eV shower. The zenith angle of the
shower is 13$^{\circ}$ indicating a young shower. Each panel shows the 
trace measured in  an Auger  water tank at the indicated distance. 
The strength of the
signal is also indicated. The start time of each trace is arbitrary. The
trace at 1680 m shows evidence of individual muons.}
\end{figure}

\begin{figure}
\centerline{\hbox{\psfig{figure=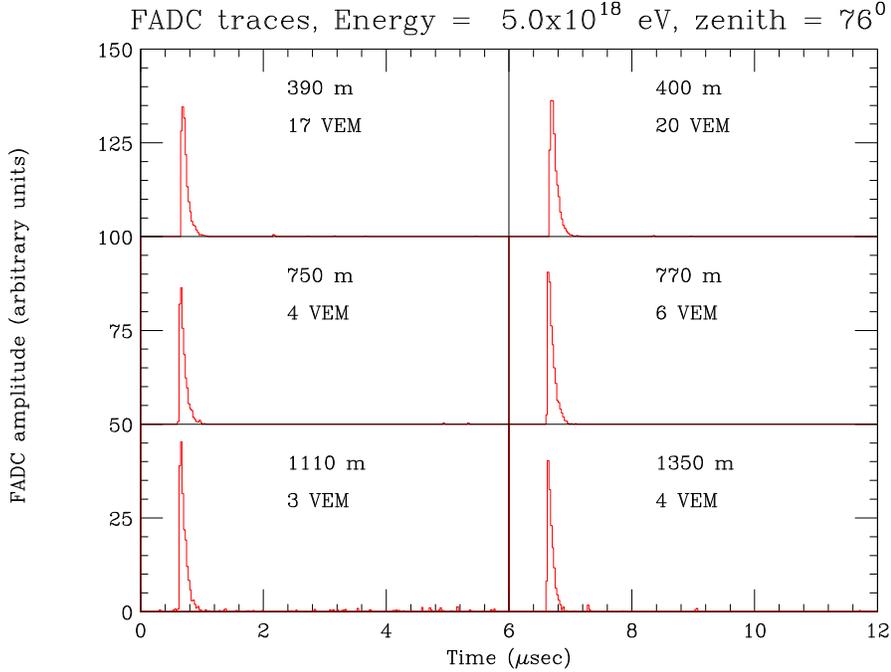,height=3.5in,angle=90}}}
\caption{FADC traces for a $\sim$5x10$^{18}$ shower. The zenith angle of the
shower is 76$^{\circ}$ indicating an old shower. Each panel shows the
trace measured in  an Auger  water tank at the indicated distance. 
The strength of the
signal is also indicated. The start time of each trace is arbitrary. }
\end{figure}

There is an additional property of showers that has yet to be fully exploited
\cite {Linsley3}.
This is the fact that the shower particles arrive spread out in time as
they are detected away from the shower axis. The spread in time of the particles
beyond 1.0 km is measured in $\mu$-sec and is
 approximately proportional to the the distance to the shower
axis. The energy deposited by the shower particles  
in the Auger water detectors is recorded in individual time slots of 25 ns
width. Fast analog to digital convertors (FADCs) are used for this purpose. 
This energy deposit is measured in
units of VEM, (vertical equivalent muons, the signal produced by a single
muon traversing the center of the tank). The distribution of arrival times
for a shower with zenith angle of 13$^{\circ}$ is shown in Figure 18. This 
shower of about 10$^{19}$ eV is called  ``young'' as its maximum occurs 
at about 750 gm/cm$^2$ close to the ground level of 850 gm/cm$^2$. By contrast
large zenith angle showers present a strikingly different pattern as shown in
Figure 19. Here the particles arrive within a fraction of a $\mu$-sec for a
shower with zenith angle 76$^{\circ}$. This shower is called ``old'' as it
is observed after having passed through 3500 gm/cm$^2$ before reaching the 
ground. These old showers consist entirely of muons (with $\sim$ 20$\%$ 
electromagnetic fuzz clinging to them)  and have a very sharp shower front.
The electromagnetic part of the shower has died out. One can note as well
that the young shower has a much steeper lateral distribution, a signal ratio
of 140 for a distance ratio of 3.6. For the old shower the signal ratio is
only 4 for a distance ratio of 3.5.

\begin{figure}
\centerline{\hbox{\psfig{figure=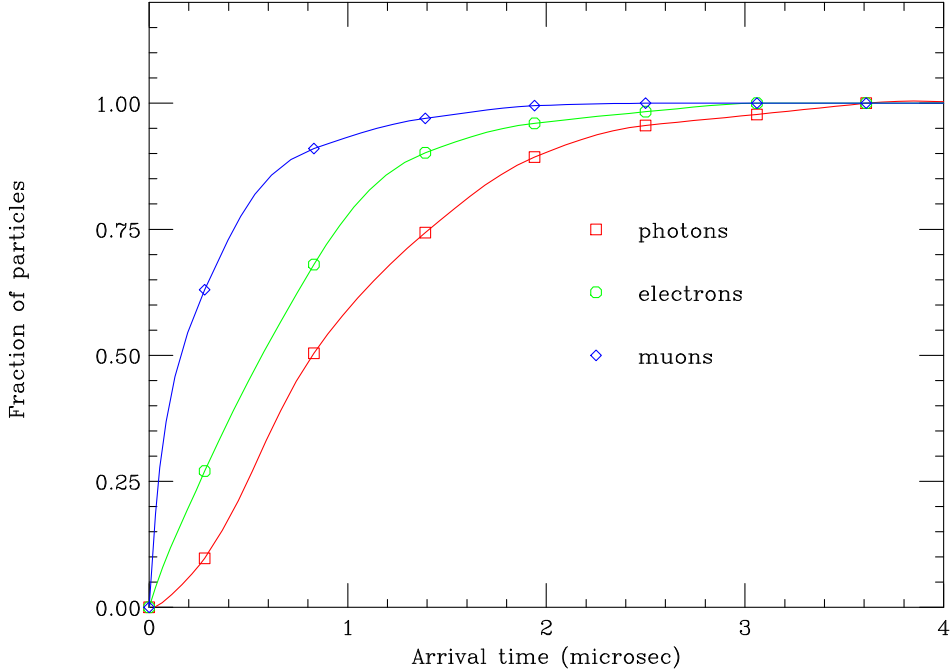,height=3.5in,angle=90}}}
\caption{Integral arrival time for the components of a young shower of
10$^{19}$ eV. The distance to the shower axis is 1 km.}
\end{figure}

The shape of the trace for young showers contains information about the 
relative amount of muons and electromagnetic particles in the shower. 
Figure 20 shows
a simulation of the integrated arrival times for muons, electrons and
photons for a typical young shower initiated by a proton of 10$^{19}$ eV.
Geometrical
arguments lead to the conclusion that the more distant the shower maximum
is from the ground, the tighter the spread of particles. Iron primaries 
have a more distant shower maximum and more muons, which will result
in a sharper pulse. Analysis of the shapes of FADC traces can help 
reveal statistically the nature of the primary particle.

\section{The two large experiments}

There are two cosmic ray experiments that now dominate the experimental
scene. The most mature of the two in its analysis and the best documented is 
AGASA (Akeno Giant Air Shower Array) \cite{agasa2}. AGASA operated for 
12 years and ceased operation
January 4, 2004. The High Resolution Fly's Eye (HiRes) has been operating 
for a few years and will continue through 2005. It is a stereo fluorescence 
detector and the analysis is at present less mature than AGASA. 

\subsection{AGASA}

The design of the AGASA was made in the late 1980's before the availability
inexpensive FADCs. The particle densities were measured 
logarithmically.
The calibration was based on measurements of single muons in the cosmic rays. A
number of technical criticisms accumulated over the years concerning the
response of the electronics and the sensitivity of scintillators to
delayed slow neutrons associated with the shower. The reference \cite{agasa2} 
addresses in a satisfactory way these criticisms, and the reader is encouraged 
to consult this paper.

AGASA was a surface array consisting of 111
scintillator detectors spread over an area of 100 km$^2$. The scintillators are
2.2 m$^2$ in area and 5 cm thick. In addition there are 27 muon detectors
in the southern part of the array. These are shielded particle detectors 
ranging in area from 2.8 to 10 m$^2$. 
The mean atmospheric depth is 920 gm/cm$^2$.
The most recent report was based on an exposure of 5.3 x 10$^{16}$ 
meter$^2$-sec-sr for showers with zenith angles less than 45$^{\circ}$
\cite{agasa3}. 

The  shower energy is 
proportional to the density of particles recorded at 600 meters. The
scintillator measures almost exclusively the electron component of the
shower.
 The sensitivity to the photon component is less 
than 10$\%$. The contribution of muons to the signal at 600 meters is about 
10-14$\%$, referring to Figure 13. The signal at 600 meters is insensitive
to the primary composition and on average independent of the position of
shower maximum. It was Hillas \cite{Hillas2} who first pointed out that the 
density at 600 meters is a good energy parameter. The distribution of shower 
particles is at any depth 
cylindrically symmetric in a plane perpendicular to the shower axis. However
the plane of the ground is only concident with a shower plane for vertical 
showers. The signals are placed in an effective shower plane by
plotting their perpendicular distances to the shower axis. For inclined 
showers an azimuthal
asymmetry is generated about the shower axis.
For a shower of inclination of 30$^{\circ}$ the atmospheric depth 
at 600 meters from the shower axis is
modulated  by $\pm$40 gm/cm$^2$ and at 45$^{\circ}$ by 
$\pm$70 gm/cm$^2$. These azimuthal variations are ignored. As the showers are
near maximum development and as in most events a number of azimuths are 
sampled, the modulation usually averages out.

\begin{figure}
\centerline{\hbox{\psfig{figure=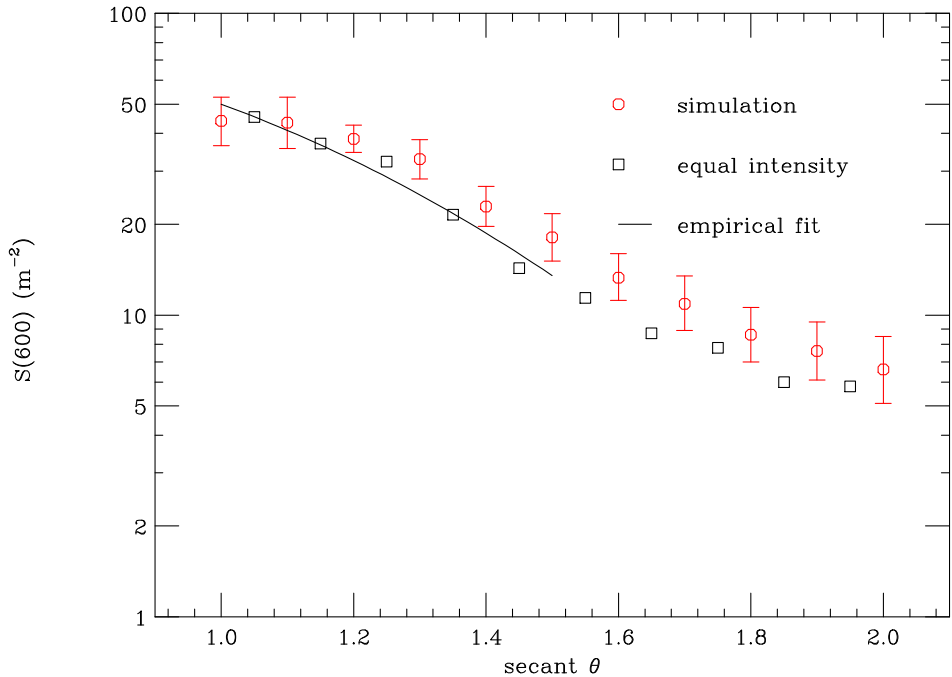,height=3.5in,angle=90}}}
\caption{Dependence of the particle density at 600 meters for the AGASA
array on zenith angle for showers of 10$^{19}$ eV. Shown are the 
experimentally measured points using the equal intensity method. The
solid line is the empirical fit used to obtain S$_0$(600), the equivalent
value for a vertical shower. Also shown are the points produced by simulation.
The absolute values are normalized by the simulation.}
\end{figure}

What is important however is the general attenuation of the entire shower,
since the atmospheric depth increases with inclination. This attenuation, 
at energies where there are many showers, can be directly measured. For a
given energy the true flux of cosmic rays should be independent of the
zenith angle. By selecting showers in bins of zenith angle and
demanding equal rates, one can relate the density of particles at 600 m
with that for vertical showers, and thus determine for a given energy the
density as a function of zenith angle. This ``equal intensity'' method 
can be checked by simulations. The measured relative response and the
results of simulations for the AGASA array are shown in Figure 21. 
Also plotted is the empirical relation used to convert the observed density 
to what its value would be for a vertical shower. That density corrected
to vertical is proportional to the energy.

\begin{figure}
\centerline{\hbox{\psfig{figure=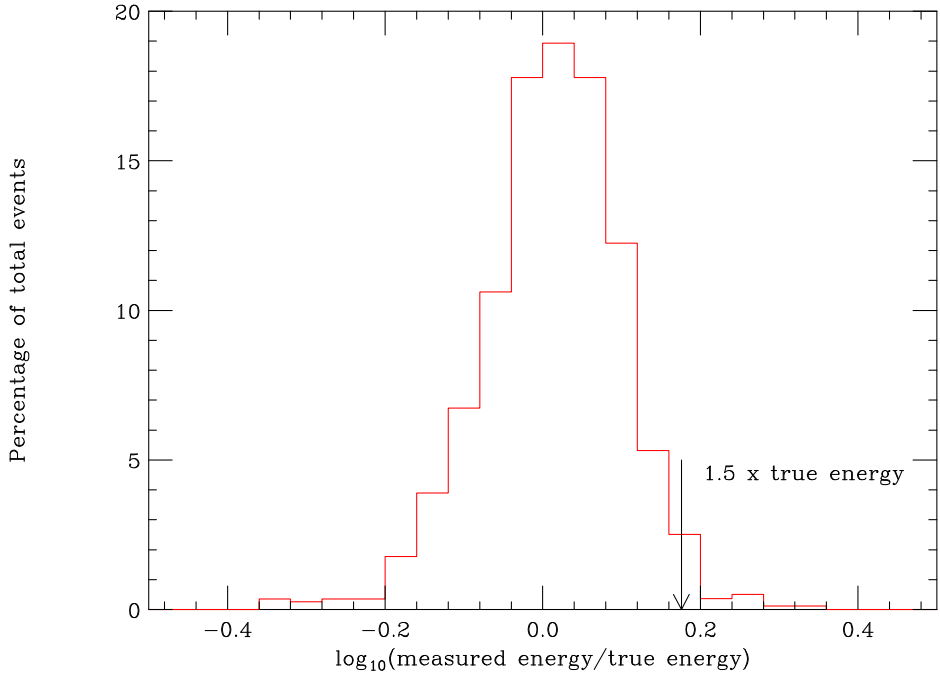,height=3.5in,angle=90}}}
\caption{The AGASA energy resolution for showers of 10$^{20}$ eV.}
\end{figure}

The errors in energy reconstruction are calculated by simulation, where all the
input is based on extensive experimental measurements. The resolution is 
presented in logarithmic units which is natural for the typical scale of 
errors in cosmic
ray experiments. The natural unit here is 0.1.  In Figure 22 we reproduce 
the plot of the expected energy
fluctuation for showers of 10$^{20}$ eV for the AGASA experiment. The
curve is asymmetric with a tail on the low energy side. The full
width at half maximum (FWHM) is 0.2 corresponding to an RMS error of 
$\pm$ 21$\%$.
More important is that probability of an upward  fluctuation to $\geq$ 1.5 times
the true energy is 2.8$\%$. Figures for showers of 3 x 10$^{19}$ eV are 0.25
FWHM and 2.3$\%$ for an upward fluctuation beyond 1.5. With this resolution
a GZK cutoff cannot be transformed into an excess of post-GZK events.

\subsection{HiRes}
 
The HiRes experiment is an extension of the original pioneering Fly's
Eye experiment \cite{flyseye1}. The technique is to observe the yield
of 300-400 nanometer photons produced by the fluorescence of charged
particles passing through the nitrogen of the atmosphere. The fluorescence
yield per unit track length is isotropic and
roughly independent of the atmospheric pressure.
One would expect the fluorescence emission to be proportional to dE/dx
which would produce more photons per unit track length at higher pressure.
The competition between deexcitation by collisions which increases with
pressure and radiation deexcitation produces a fluorescence yield per
unit track length that is nearly independent of pressure. The technique
measures most directly the number of charged particles as the shower
develops longitudinally. To get the corresponding energy deposit, one needs 
to know the  density of the atmosphere corresponding the point in 
the sky where the fluorescence light is emitted. Thus, it is crucial to know
independently the density of the atmosphere as a function of altitude.
It is said that the fluorescence technique is ``calorimetric'', but a
purist might disagree.

The HiRes experiment presently operating consists of
two stations separated by 12.6 km. One station (HiRes I) has 21 mirrors with
a view of the sky of 360$^{\circ}$ in azimuth and 3$^{\circ}$ to 17$^{\circ}$
in elevation. 
The second station (HiRes II) has a view of 360$^{\circ}$ in azimuth and 
3$^{\circ}$ to 31$^{\circ}$ in elevation. The sky is divided into pixels
of 1$^{\circ}$ by 1$^{\circ}$ by clusters of photomultipliers at the focus of
each 3.8 m$^2$ mirror.

\begin{figure}
\centerline{\hbox{\psfig{figure=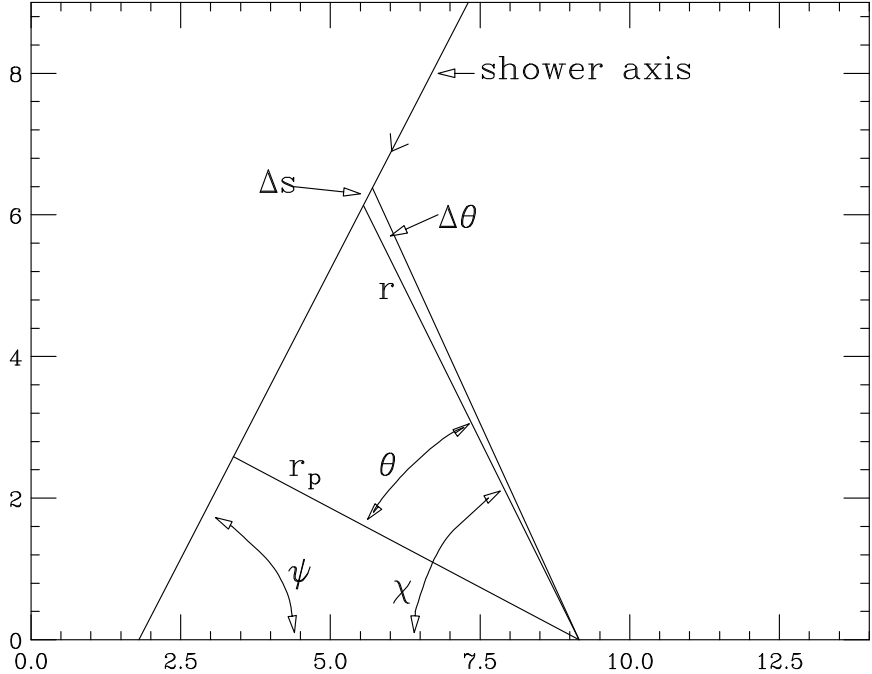,height=3.5in,angle=90}}}
\caption{Sketch for the purpose of calculating the fluorescence yield.
The indicated coordinates are proportional to distance}
\end{figure}

\begin{figure}
\centerline{\hbox{\psfig{figure=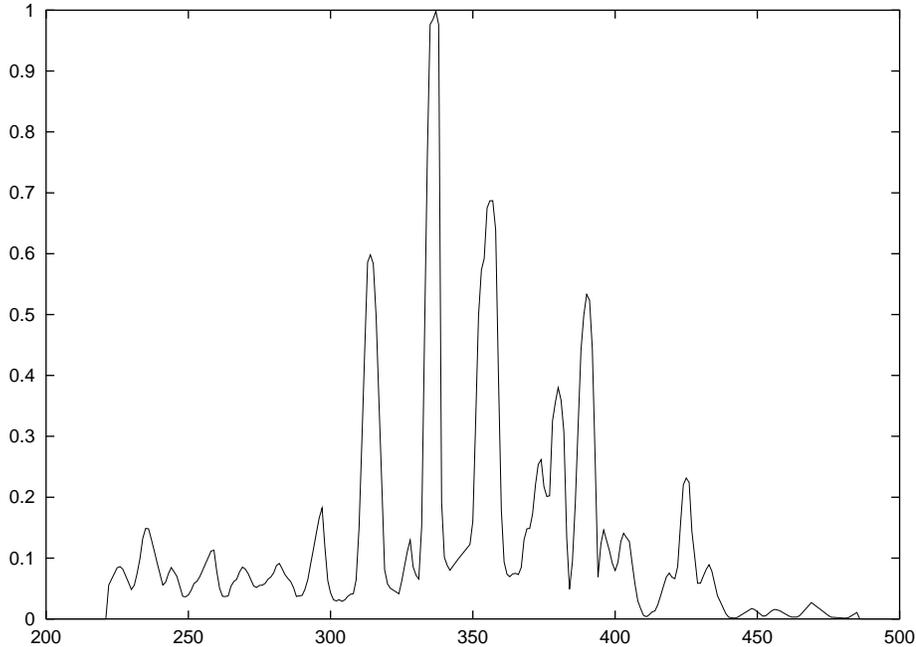,height=3.5in,angle=-90}}}
\caption{The spectrum of nitrogen fluorescence light. The abcissa is the
wavelength in nanometers. The ordinate is the relative intensity.}
\end{figure}

In Figure 23 we present the geometry for the calculation of the fluorescence
light produced by a shower. The axis of the shower and the position of the
detector define the shower-detector plane.
To a good approximation the light can be assumed to
be emitted from a line source. In an element of length $\Delta$$s$, $N_e$ 
charged particles will emit $\eta$$N_e$ photons in the range 300-400 nm,
where $\eta$ is the fluorescence yield per charged particle per meter.
The fluorescence spectrum is shown in Figure 24. This is the spectrum
measured by Bunner \cite{Bunner}.
The distance from the line element to the fluorescence detector is $r$. The 
perpendicular distance from the shower axis to the detector is $r_p$.
The number of photoelectrons received at the detector is given by:

\[ p_e =\left[(\Delta s \eta) \epsilon \left( \frac{A}{4 \pi r^2}\right) \delta 
\right] N_e . \]

Here $A$ is the area of the light collecting mirror in m$^2$, $\epsilon$ is
the product of the reflectivity of the mirror, the transmission coefficient
of the near UV filter normally used and the photocathode efficiency, $\eta$
is the fluorescence yield in photons/meter, 
and $\delta$ is the transmission of 
the fluorescence light over the distance r. With the auxiliary relations
$\Delta$$s$ = $r$$\Delta$$\theta$/cos($\theta$) and 
$r$ = $r_p$/cos($\theta$) one
finds yield of photoelectrons to be:

\[ \frac{p_e}{A} = \frac{\eta \epsilon \Delta \theta}{4 \pi} 
\frac{\delta}{r_p} N_e .   \]

It is instructive to put in some numbers. Roughly $\eta$ $\sim$ 4.5 UV
photons/meter and $\epsilon$ $\sim$ 0.16. Taking $\Delta$ $\theta$ =
0.0174 (1$^{\circ}$) one finds:

\[ \frac{p_e}{m^2 deg} \sim 10^{-6} \frac{\delta}{r_p} N_e , \]

\noindent where $r_p$ is expressed in kilometers.

Typically at the shower maximum there are about 7 x 10$^9$ charged particles
for a shower of 10$^{19}$ eV. Typical attenuation lengths are about 10 km.
Thus one expects for r$_p$ = 20 km a fluorescence signal of about 50 p$_e$/m$^2$ for a 
pixel of 1 degree$^2$. Note that $\delta$ is not a constant but depends
on the atmospheric transmission between the shower axis and the detector 
for each pixel.

The number of photoelectrons  per square meter of mirror area and
per degree of trajectory in the sky, recorded at the entrance of the
detector, is a good place to match the sensitivity of the fluorescence
detector to the external quantities, which are the geometry of the
shower and the attenuation of the atmosphere. The absolute
calibration of the fluorescence detector is a delicate matter and we will
not go into the details here \cite{matthews1}.

\begin{figure}
\centerline{\hbox{\psfig{figure=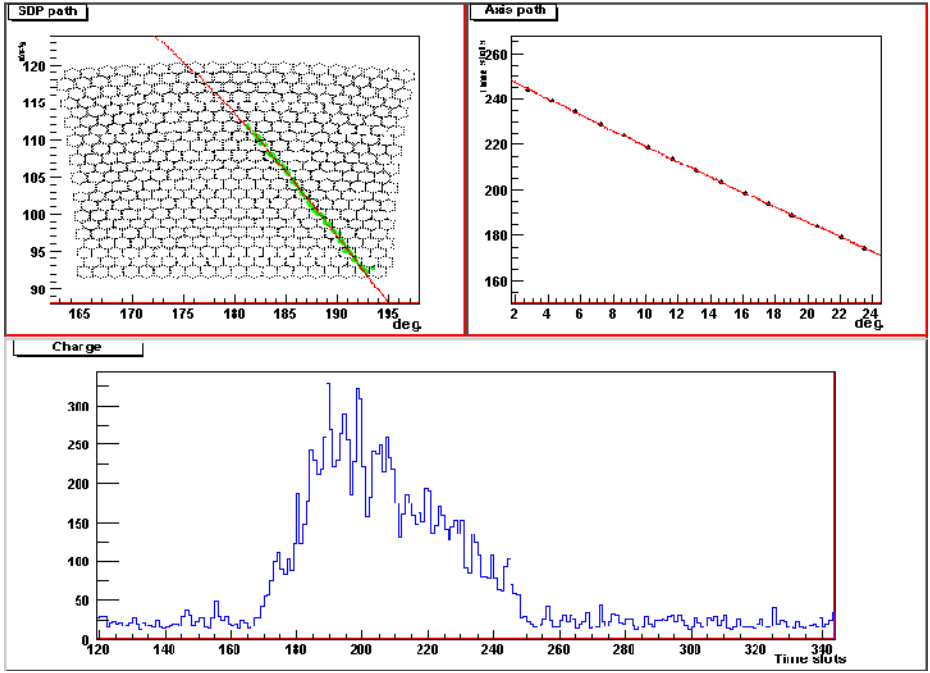,height=3.5in,angle=-90}}}
\caption{Example of the measurements made on a shower by a fluorescence
detector.}
\end{figure}

As the shower develops Cherenkov light is produced. The number of Cherenkov 
photons is much greater than the number of fluorescence photons, but 
the former are very closely 
collimated along the shower axis. Nevertheless significant Cherenkov light
can accompany the fluorescence signal. If the shower axis is directed at angles
less than 30$^{\circ}$ to the 
the fluorescence telescope axis, direct Cherenkov light will be 
detected. As the shower develops, the Cherenkov light accumulates along the
shower axis, and this light is scattered into the
field of view of the fluorescence telescope. There is also the night
sky background which is ever-present. The night sky background is typically
40 photons per m$^2$ per $\mu$sec per deg$^2$ of sky. With a mirror area of
3.8 m$^2$  the mean night sky rate is roughly 23 p$_e$ per $\mu$sec for each
pixel. The Poisson fluctuations of this background rate modulate the 
fluorescence signal. 

The measurements made by a fluorescence detector are shown in Figure
25. These are data from one of the Auger detectors, but are typical of the 
HiRes detector as well. In the upper left panel is the trace of the
shower trajectory on the focal plane of the mirror. This information
combined with the pointing direction of each pixel is used to construct the
shower detector plane (SDP). The normal to this plane can be determined to 
an accuracy of a few tenths of a degree. In general the SDP can be strongly 
tilted with
respect to the ground. In these cases the angular sweep of the fluorescence
light can be much larger than the elevation aperture of the telescope.
In the upper right panel is a plot of the time
of arrival of the light in each pixel vs its angle $\chi_i$. These are
the data needed to determine the location of the shower axis within the
SDP. The lower panel is a plot of the light observed as a function of time
in bins of 100 nsec (a FADC at 10 MHz). The night sky noise and the pedestals
are evident before and after the fluorescence signal.

\begin{figure}
\centerline{\hbox{\psfig{figure=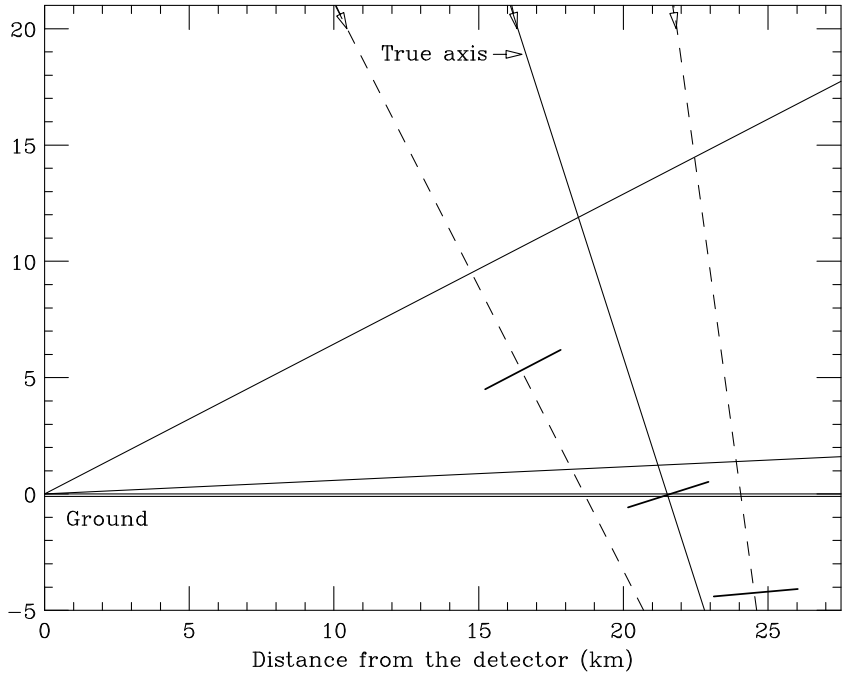,height=3.5in,angle=90}}}
\caption{Plot of a true shower axis and two other shower axes that can also
fit the angle-time data. The ambiguity is resolved by measurement of the
time of arrival of the shower front on the ground (see discussion of hybrid
method in text). If the axis were closer
to the eye, the shower front would need to be delayed (not yet at ground 
level) since the light takes less time to get from the axis to the eye.
Supposing the axis to be further from the eye would require the shower front 
to arrive ahead of the true front.}
\end{figure}

The angular velocity of the image of a vertical shower across the 
fluorescence camera is roughly
about 1$^{\circ}$/$\mu$sec at a distance of 20 km. It can be much faster 
if the shower
is aimed towards the fluoresence telescope and much slower if it is aimed
away from the telescope. HiRes I collects the signal from each pixel with
sample-and-hold electronics with $\sim$ 5$\mu$sec integration time. HiRes II
collects the signal with a 10 MHz FADC.

The location of the shower
axis within the SDP  is much less well determined and depends on the
relative time of arrival of the fluorescence signals as a function of pixel
angle. An individual skilled in trigonometry can show from Figure 23 that:

\[ t_i - t_0 = \frac{r_p}{c} tan \left( \frac{\pi - \psi -\chi_i}{2}\right) , \]

\noindent where $t_i$ is the arrival time of the fluorescence 
light at the $i$th
pixel at the angle $\chi_i$, $t_0$ is the time of emission of light at
the point defined by r$_p$,
and $c$, the velocity of light.
The quantities $\psi$ and $r_p$ define the position and angle of the
shower axis in the  SDP and hence the shower geometry is
completely determined. The angle $\pi-\psi-\chi$ is the angle between the
shower axis and the trajectory of the light ray to the detector. For
short segments and/or segments roughly perpendicular to the direction of
observation, many combinations of $r_p$ and $\psi$ can satisfy the angle-time
relation. An example of this for a SDP perpendicular to the ground
is shown in Figure 26. The true shower axis
is the solid one, but axes at $\pm$ 10$^{\circ}$ to the true one
produce only a slightly different angle time relation. Information that
breaks the degeneracy of many possible sets of $\psi$ and $r_p$ depends
on a measurable curvature in the angle time relation. For the example
shown in Figure 26 the sagittas (a measure of curvature) deviate by only 
$\pm0.4\mu$sec for the dashed trajectories. 
Consequently a single fluorescence detector (monocular) has a very asymmetric 
angular resolution.

\begin{figure}
\centerline{\hbox{\psfig{figure=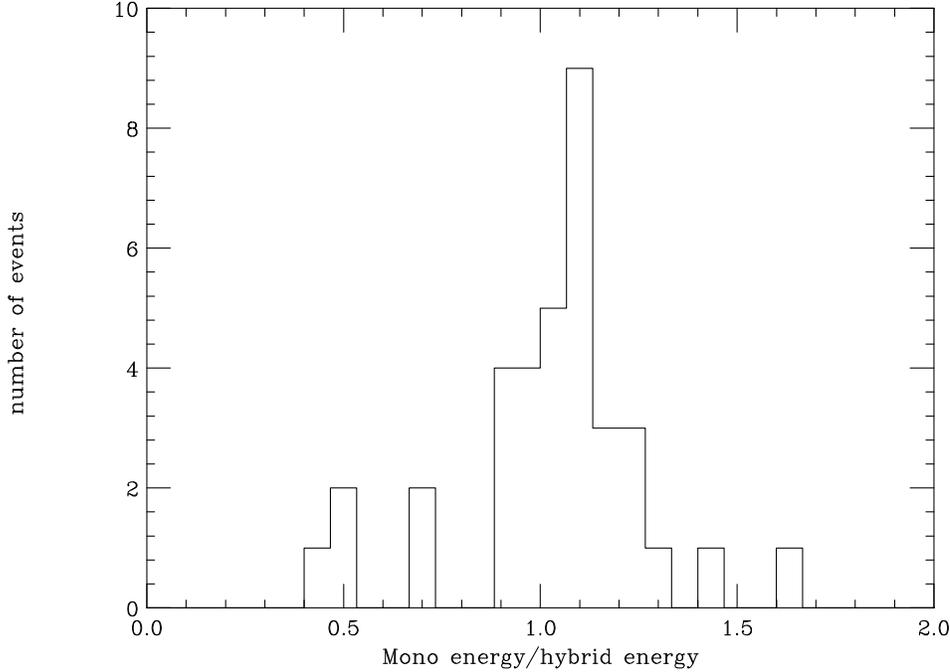,height=3.5in,angle=90}}}
\caption{Ratio of energy by monocular reconstruction to energy by
hybrid reconstruction for showers observed in the hybrid mode by
the Auger Observatory. The energy of the showers is between 10$^{18}$
and 10$^{19}$ eV.}
\end{figure}

The conversion of the observed light pattern in the FADC to
number of particles as a function of atmospheric depth in gm/cm$^2$ 
depends on the zenith angle of the shower axis and a knowledge of the density
of the atmosphere as a function of height. A poor measurement of the
angle $\psi$ and $r_p$ will result in a poor energy measurement. To avoid
this problem the original Fly's Eye experiment \cite{flyseye2} and the
HiRes experiment chose to view the showers in stereo. Then the intersection
of the two SDPs gives an accurate measure of the shower axis. 

A second method which requires only a single fluorescence telescope uses the 
measurement of the relative time of arrival of the fluorescence light with the 
time of arrival of shower particles in one or more surface detectors. As
shown in Figure 26, a measurement of the arrival time of the shower
particles selects the true trajectory.
This hybrid technique \cite {sommers1} is employed by the Auger Observatory. 
The precision is as good or better than a stereo arrangement. 

The Auger Observatory with a small number of surface detectors was operated 
for a few months in a hybrid mode. A coincidence between a fluorescence 
detector and even a single surface detector gives a precision for the 
intersection of the shower axis with the ground of better than 100 meters. 
In Figure 27 we show the ratio of the energy by mono reconstruction to hybrid 
reconstruction which is a measure of the accuracy of the mono reconstruction. 
While the circumstances are not exactly the same as with HiRes, one gains some 
appreciation for the accuracy of a mono analysis which is not so bad 
($\pm$ 25$\%$).

\begin{figure}
\centerline{\hbox{\psfig{figure=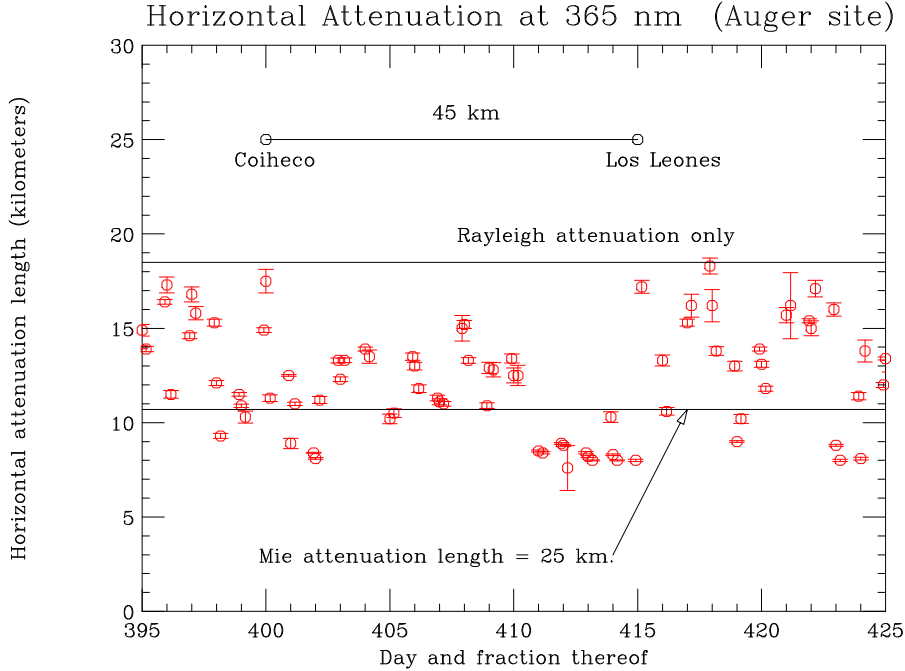,height=3.5in,angle=90}}}
\caption{The horizontal attenuation length at the Auger Observatory
site, 1400 meters elevation.}
\end{figure}

The  FADC light trace in Figure 25 must be converted to the number
of charged particles as a function of atmospheric depth. This conversion
involves correction for the attenuation in the atmosphere. There are two
components to this correction, the Rayleigh scattering in the air, and
the scattering of by aerosols in the air (Mie scattering). The latter can be
highly variable and must be measured. If the vertical structure of the
atmosphere is known, the effects of Rayleigh scattering can be evaluated.
A discussion of these most important corrections are beyond the scope of this
article. A very clear discussion of the atmospheric corrections can be found in
the reference \cite{matthews2}. While this paper refers to the atmospheric
monitoring program of the Auger Observatory, it shares many of the 
features of the HiRes program
and has greatly benefited from collaboration with the HiRes group. The Rayleigh
attenuation length at the elevation of the HiRes site is $\sim$18 km. It
lengthens with altitude as the density of the atmosphere decreases as
e$^{-(h/7.5)}$ where h is the altitude in km. Typical
values of the aerosol absorption length are about 25 km, but are highly
variable. The scale height for the aerosols is usually much smaller than
the atmosphere being,
$\sim$ 1.2 km. For proper correction extensive atmospheric monitoring
is employed using LIDAR and horizontal attenuation monitors. The
horizontal attenuation length of the atmosphere at the detector level shows in
dramatic fashion the variation of the atmosphere. The horizontal attenuation
length at the Auger site is plotted for a month period in Figure 28. There
are occasions where the attenuation length shows the complete absence of
aerosols.

\begin{figure}
\centerline{\hbox{\psfig{figure=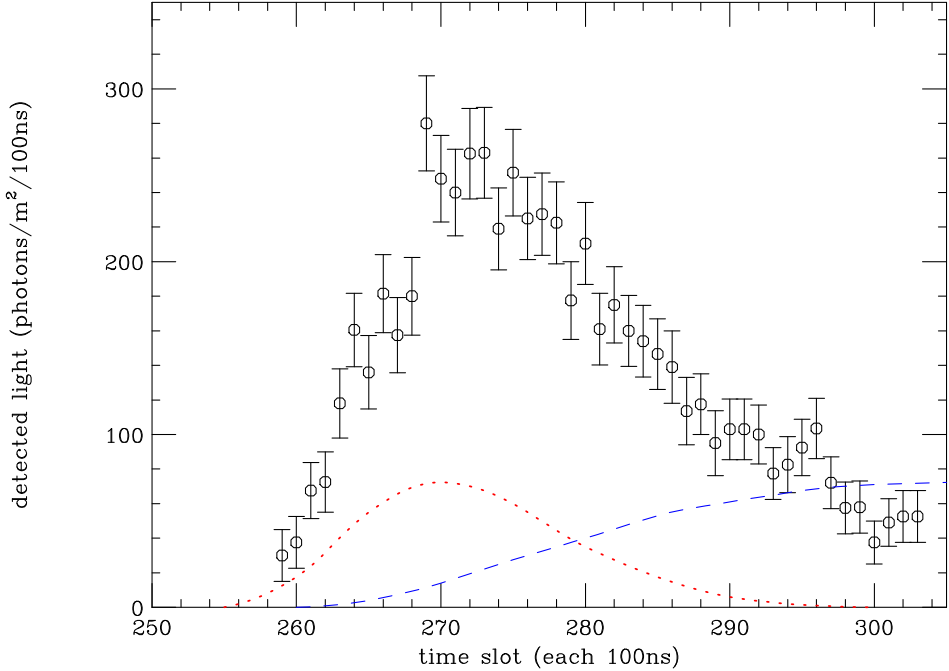,height=3.5in,angle=90}}}
\caption{A typical light curve vs time
measured by one of the Auger fluorescence telescopes. The points with errors
are the observed photons/m$^2$/100ns entering the telescope.
Shown is the total 
signal and the contributions to it by Cherenkov light. The direction of this
shower was
chosen to show both direct Cherenkov light (dots) and scattered Cherenkov
light (dashes).} 

\end{figure}

The current analysis of the HiRes data has been made with average values for
the aerosol scattering and the aerosol scale height. This procedure is
sufficient for analysis at no better than a 25$\%$ level. The ultimate analysis
will certainly take into account the variability of the atmosphere.
Typical attenuation lengths are about 15 km. At 20 km this corresponds
to a correction of a factor 4 and at 30 km a factor 7. As the highest-energy
events require the largest aperture, they will be the most distant
and will require the largest correction. The attenuation correction is of course
not a single number, but depends on the geometrical path of the light to each
pixel. The depth of shower maximum of the light curve depends on the 
attenuation correction.
Finer details can be imagined. The fluorescence yield consists of discrete
bands ranging from 310 to 390 nm.  The Rayleigh attenuation length, which
varies as $\lambda^{4}$, changes by nearly a factor of 5 over this range.
Thus, it is important to know not only the overall fluorescence yield
but, also its individual components very accurately.

A typical light curve observed by an Auger fluorescence telescope is shown 
in Figure 29. The light curve is a mixture of fluorescence light and
Cherenkov light. In this particular case there are contributions for both
direct Cherenkov light which contaminates the early part of the light curve 
and scattered Cherenkov light which is important near the tail of the signal.
An iterative technique is used as the strength of the Cherenkov signal
depends on the $N_e$ which is derived from the fluorescence signal itself.
The geometry of the shower axis with respect to the telescope must be
reconstructed to determine the Cherenkov subtraction. 
The energy is determined from the ionization loss of the charged particles.
In addition some 5 to 10$\%$ of energy is not visible due to muons and
neutrinos.

The fluorescence technique is a very beautiful one. However it needs
enormous attention to details to exploit its ultimate precision. These
details include
absolute calibration of the optical system, gain
monitoring, atmospheric monitoring, absolute knowledge of the nitrogen
fluorescence efficiency, and the vertical profile of the atmosphere which 
varies between summer and winter. The detector can only operate on dark 
moonless nights which
limits its duty cycle to $\sim 10 \%$.
The viewing time is longer in the winter.  Many years are required 
to achieve uniform exposure in right ascension. The surface array is 
robust and can
operate in all weather with $\sim 100 \%$ duty cycle. Nature provides
a constant calibration in the form of an abundance of single cosmic ray muons.
However without an independent means of calibration simulations are required to
establish the energy scale.

In metaphorical terms the fluorescence technique resembles a beautiful 
prima donna who needs constant pampering. Then she will sing with such 
beauty that shivers run up and down your spine. By contrast the surface 
array technique reminds one of a chanteuse in a smoky bar who sings with the
same passion, no matter how she feels or how she  is treated.

\section {Experimental results}
\subsection{Energy spectrum}

The HiRes and AGASA detectors are located at approximately the same
latitude, 40.3$^{\circ}$N and 35.5$^{\circ}$N respectively. Thus they
look at the same region of sky and are expected to observe the same
spectrum of cosmic rays. This would not necessarily be the case for a 
comparison with a spectrum from the southern hemisphere. The data presented
by the AGASA array has been accumulated over 12 years \cite{agasa3}. 
Their analysis is very mature and they have responded in detail to 
to a number of thoughtful criticisms \cite{agasa2}. In this paper the
AGASA group estimates the systematic error of their energy measurement to be
$\pm$ 18$\%$. The aperture of AGASA is 
flat above 10$^{19}$ eV as its calculation is insensitive to the exact 
details of the trigger since many detectors are hit. The aperture 
decreases as the
energy drops below 10$^{19}$ eV. The efficiency to record a shower
then depends on the position of the shower on the array and the criteria
for a scintillator to trigger. The calculation of the aperture is
critical in the conversion of the number of events per energy bin to
a flux. When the array is not fully efficient this calculation is sensitive
to precise understanding of the individual detector and the details of the
lateral distribution.

\begin{figure}
\centerline{\hbox{\psfig{figure=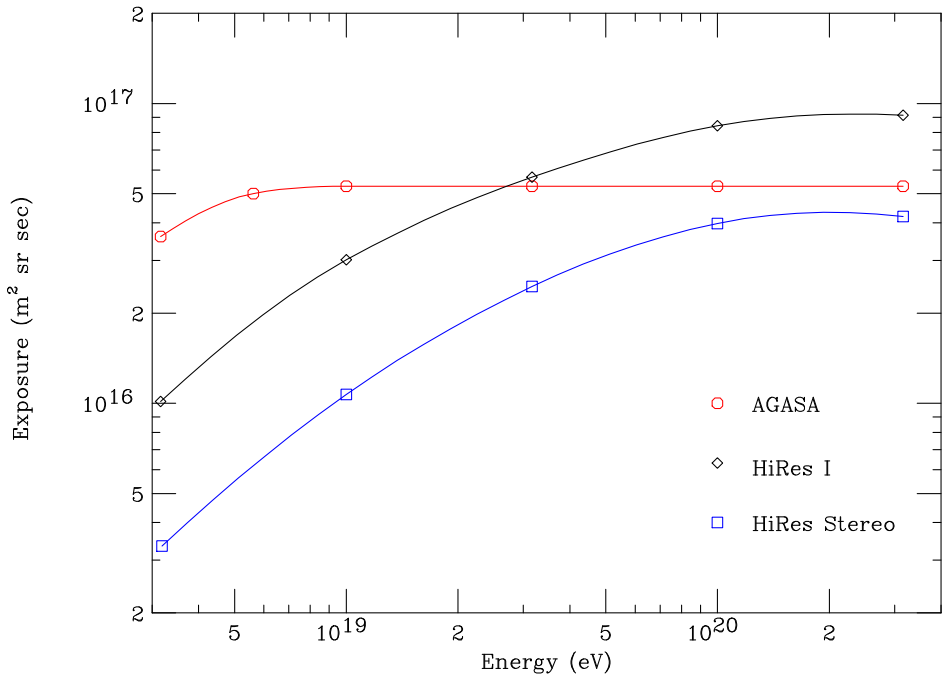,height=3.5in,angle=90}}}
\caption{The exposure of the AGASA experiment, the HiRes I monocular
experiment, and the HiRes stereo experiment as reported at the 2003 ICRC.}  
\end{figure}

\begin{figure}
\centerline{\hbox{\psfig{figure=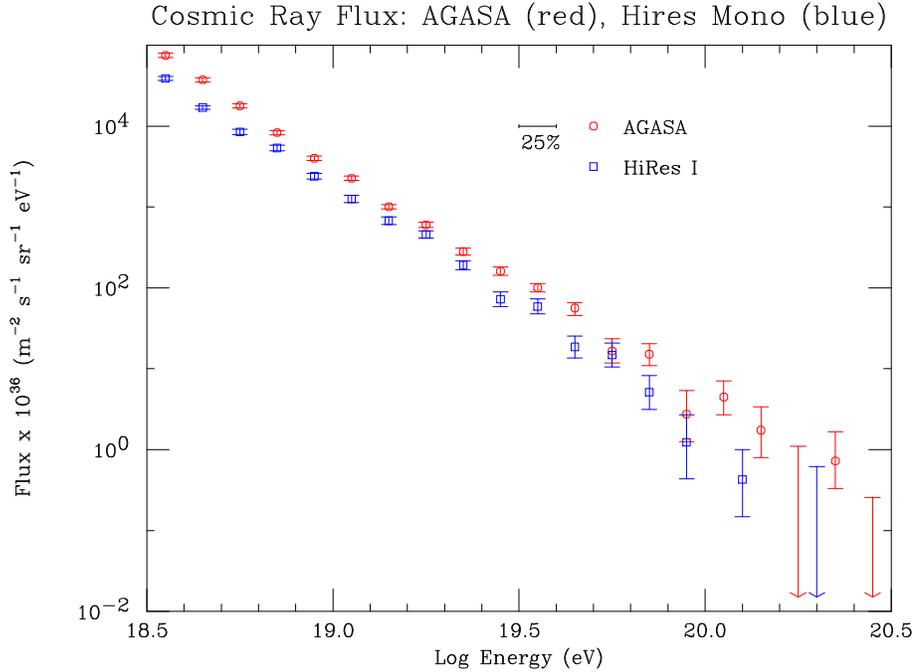,height=3.5in,angle=90}}}
\caption{Plot of the AGASA spectrum (circles) and the HiRes I
monocular spectrum (squares) for energies $\geq10^{18.5}$.}
\end{figure}

Recently the HiRes group has
reported on 5.7 years of  monocular operation from HiRes I 
\cite{abu-zayyad} with
2820 hours of data. (Periods of lack access to the Dugway site certainly 
reduced the amount of data that potentially was available.)
The group has also reported  1006 hours of
stereo data which is intrinsically more reliable \cite{springer}. 

As HiRes I has an elevation coverage from only 3$^{\circ}$
to 17$^{\circ}$ an extra constraint was required to reconstruct these
mono events. They constrained the traces to  fit  a 
Gaisser-Hillas profile with an X$_{max}$
between 680 and 900 gm/cm$^2$ to supplement the angle-time reconstruction
of the shower axis in the SDP. They estimate the RMS energy resolution to
be better than 20$\%$ above 3x10$^{19}$ eV. The data are analysed with
average values for the atmospheric conditions. It is intrinsic to the
fluorescence technique that the aperture grows with energy. 
A higher-energy shower can be seen further away. So at all energies
the calculation of the aperture depends on many details.
The authors estimate that the
total systematic error in their energy measurement is 21$\%$. 
A separate estimate of the systematic error in the aperture is not clearly
given.
The interest in this spectrum is that at 10$^{20}$ eV its exposure
exceeds that of AGASA.  The exposures for the three measurements are
plotted as a function of energy in Figure 30. The spectra for HiRes I
and AGASA are plotted in Figure 31. The two spectra in the energy range
from 10$^{19}$eV to 10$^{20}$ eV show remarkable agreement considering
an estimated systematic error of about 20$\%$ for each experiment. A
shift in energy of one or the other experiment by $25\%$ will produce good
agreement. Much has been made of the absence of events beyond 10$^{20}$
eV in the HiRes I data, notably by authors whose theories on the
origins of the highest-energy cosmic rays predict a GZK cutoff \cite{bahcall}.
I have a few comments to offer. First, the more reliable HiRes stereo data
will give a far more convincing case if the absence of events beyond
10$^{20}$ eV is to be confirmed. In addition to significantly more
stereo exposure, one expects many improvements in the analysis. Also
one cannot easily make the 11 AGASA events above 10$^{20}$ eV go away, and they
are too many to be explained as resolution spillover. Further, as
explained in the propagation part of this paper, the fluctuations in
the events found above the GZK cutoff can be very large.

\begin{figure}
\centerline{\hbox{\psfig{figure=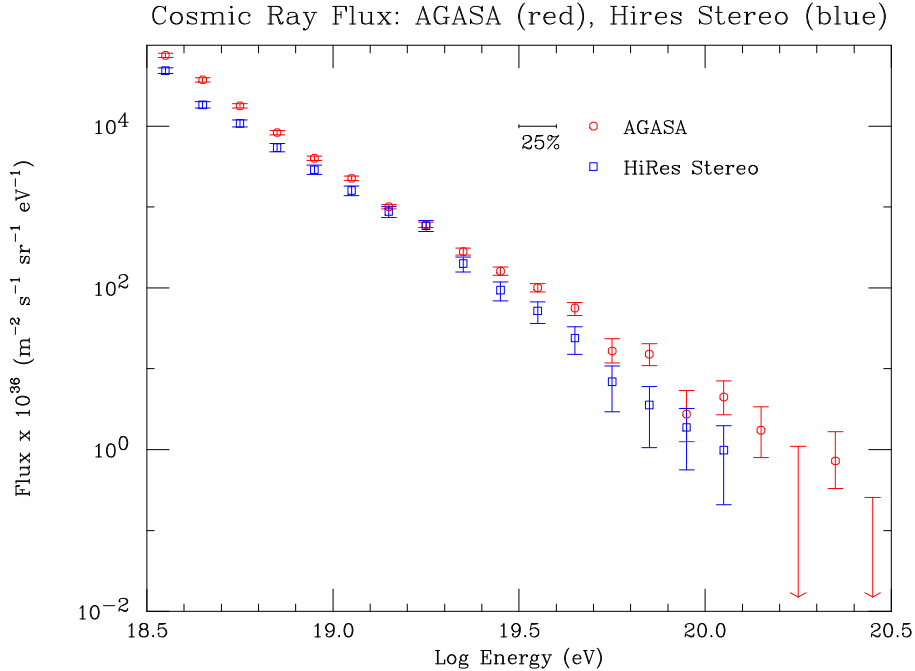,height=3.5in,angle=90}}}
\caption{The HiRes stereo spectrum as presented at the 2003 ICRC.}
\end{figure}

The HiRes stereo spectrum has been eagerly anticipated.  Its
exposure does not yet match that of AGASA. And the analysis has not yet
reached a maturity that one can ultimately expect. The average atmosphere
must be replaced by the nightly measurements 
and improvements
must be made in the energy calibration and the aperture calculation.
The experiment continues to run and ultimately it can provide the best 
measurements in the northern hemisphere for some time to come. The stereo 
spectrum as reported at the 2003 ICRC is shown in Figure 32.  For comparison 
the AGASA spectrum is replotted. The amount of HiRes stereo data in this plot 
is $\sim$ 25$\%$ of what will eventually be available.  

\subsection{Anisotropy and cluster analysis}

\begin{figure}
\centerline{\hbox{\psfig{figure=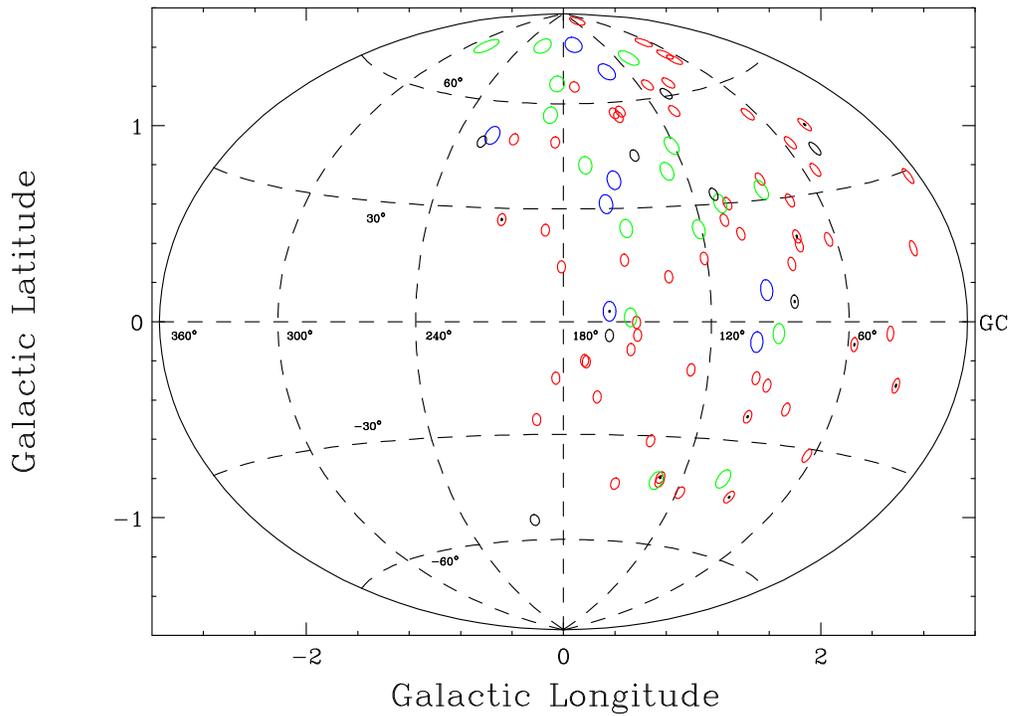,height=4.0in,angle=90}}}
\caption{Arrival directions of cosmic rays with energy $\geq$ 4 x 10$^{19}$
eV observed by AGASA, Haverah Park, Yakutsk, and Volcano Ranch. The
size of the oval boundary represents the angular resolution. A dot
within the oval indicates that the cosmic ray had an energy $\geq$
10$^{20}$eV. The small ovals come from AGASA or Volcano Ranch. The
larger ovals come from Yakutsk or Haverah Park.}
\end{figure}

The distribution of directions for cosmic rays in the northern hemisphere
is plotted in galactic coordinates in Figure 33. 
The choice of minimum energy is
4 x 10$^{19}$ eV. This energy was chosen 
by the AGASA group to make their cluster analysis. The rationale for this
choice is not clear, perhaps a balance between a sufficiently high energy
so that
the selected events have some chance to point to a source  and the desire 
to have a reasonable number of events.
There are a total of 89 events plotted, 57 events from AGASA \cite{agasa4}, 
16 from Haverah Park \cite{haverah}, 8 from Yakutsk \cite{yakutsk}, 
and 8 from Volcano Ranch \cite{volcano}. I regret that the $\sim$ 20
events from the HiRes stereo experiment were not available.

\begin{figure}
\centerline{\hbox{\psfig{figure=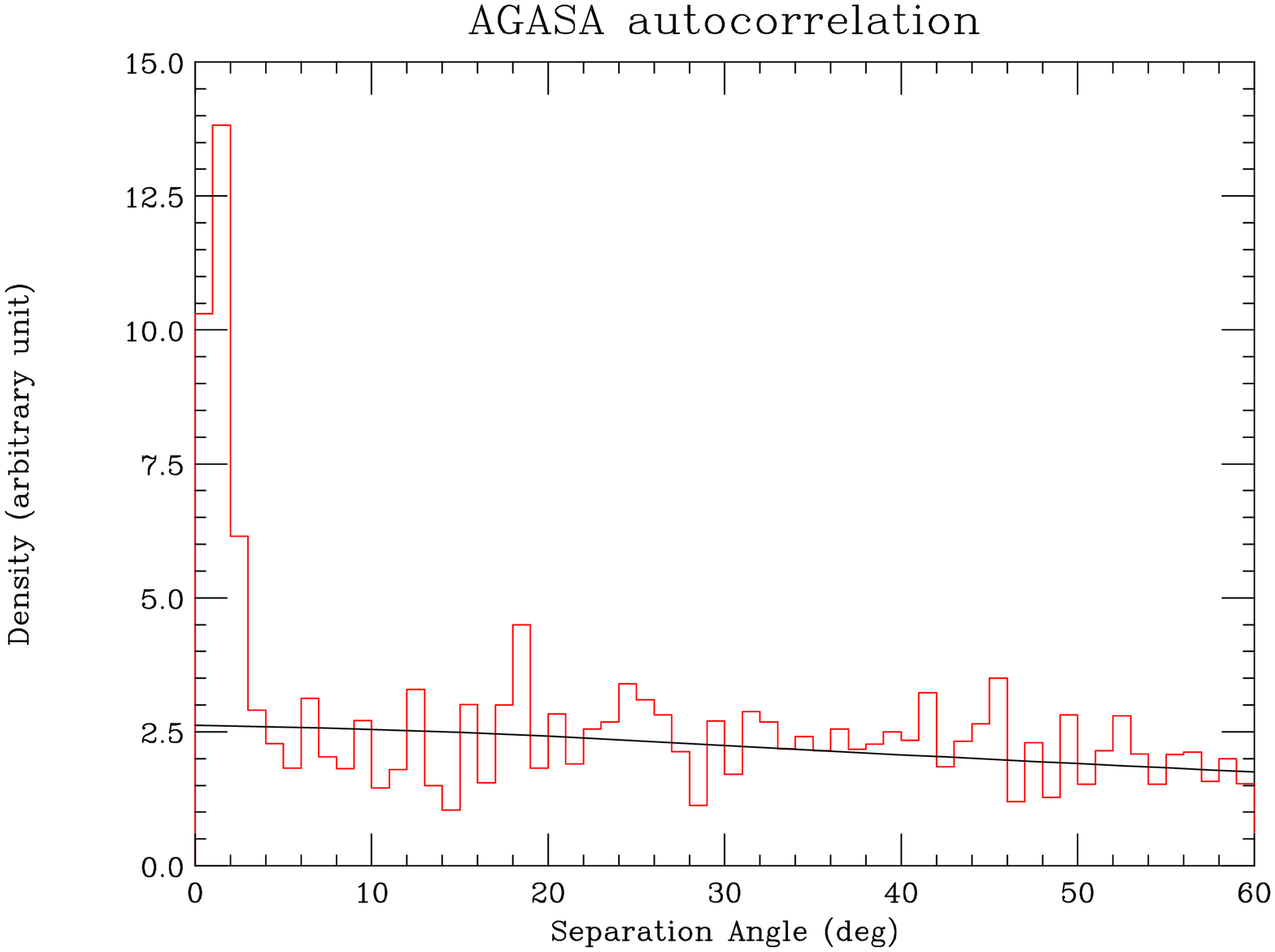,height=3.5in,angle=90}}}
\caption{Distribution of angular distances between pairs of  
59 AGASA events with energy $\geq$ 4 x 10$^{19}$ eV weighted by
the inverse of the solid angle of the angular bin. The solid line is the
distribution expected for random arrival directions.}
\end{figure}

\begin{figure}
\centerline{\hbox{\psfig{figure=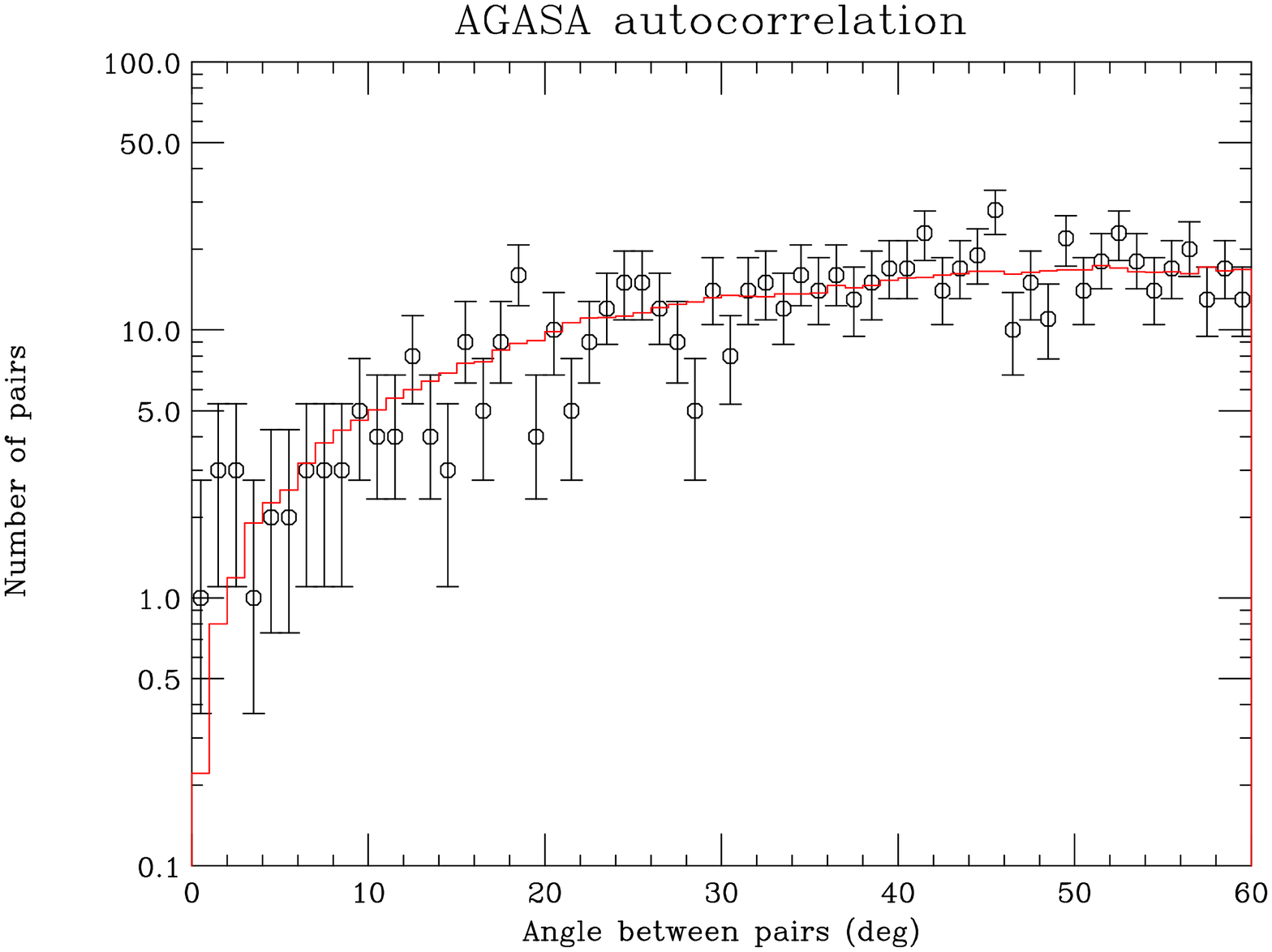,height=3.5in,angle=90}}}
\caption{Distribution of angular distances between pairs of 57 AGASA
events with energy $\geq$ 4 x 10$^{19}$ eV. The histogram is the distribution
expected for random  arrival directions. The errors
represent 68$\%$ confidence limits.}
\end{figure}

The distribution of the cosmic ray directions in the northern sky shows no 
gross anisotropy. There are 4 pairs and one triplet in the sample of AGASA
events with energy $\geq$ 4 x 10$^{19}$ eV. The pairs or triplet are defined
as two or three events whose angular separation is less than 2.5$^{\circ}$
This analysis is based on data collected through May 2000 when
the AGASA exposure was 4.0 x 10$^{16}$ m$^2$ sr sec. No public catalog
of directions has been published for more recent data although the AGASA
spectrum discused above is based on data accumulated through December 2002
\cite{agasa3}. The AGASA experiment
was terminated on Jan 4, 2004 and will have accumulated a total exposure of 
$\sim$ 6.0 x 10$^{16}$ m$^2$ sr sec. So we can expect at least 50$\%$ more
events when the cluster analysis is complete. Figure 34, which appears in 
\cite{agasa4},
is a plot of the separation angles between all pairs of cosmic rays 
in angular bins of 1$^{\circ}$. Each bin is weighted by the 
inverse of the solid angle of that bin.
There were eight pairs contributing to the 
plot, three of which are part of a triplet. However to make the eighth pair 
the limit of 4 x 10$^{19}$ eV was lowered to 3.89 x 10$^{19}$ eV to pick up an 
additional pair.  The distribution of angles for random
directions in the sky was generated from the data itself
by choosing the right ascension of each event randomly  between 0
and 360$^{\circ}$. This procedure was repeated for the data set many 
times and the resulting average distribution of angles between pairs is 
plotted as the solid line. An impressive peak results at small angles but
one has no sense of its significance as no errors are indicated.   

I have made a plot of the pair angle distribution not weighted by the
inverse of the solid angle which is shown in Figure 35. This plot includes
the 57 events from the AGASA catalog, not pulling 
in the eighth pair. 
Here the observed
number of events is plotted along with a histogram of the distribution
expected for a random right ascension. For each point the 68$\%$ confidence
limits are plotted \cite{feldman}. The information to assess the significance
is now revealed. For the first three bins 7 pairs are observed where
2.2 are expected. The Poisson probability to observe 7 or more pairs is
.0075. I will leave it to experts to evaluate the true significance, but
as experience shows, this evidence for pairs is not yet convincing. I have
made a similar analysis for pairs using the 89 events in the skymap
of Figure 33. Here for the first three bins there are 12 pairs with
an expected 6.0 for the randomized distribution. For this case the probability
to observe 12 or more pairs is .02.

The most striking aspect of the Figure 33 is the
appearance of a second triplet. I have made an  analysis similar to
the pair analysis described above. The equivalent of the angle between pairs
was the average of the three angles between the members of a triplet. Plotting
this average angle for all possible triplets yielded two triplets
within the first three angular bins just as seen by eye.  
Expected from the random distribution were 0.7 triplets. The Poisson
probability for this case is 0.10.

There is no sense to try to squeeze more significance out of these data, 
but it is instructive to play a bit. It happens that, if the threshold
is increased to 5 x 10$^{19}$ eV, the triplets remain and all pairs except
those contained in the triplets disappear. The number of events decreases
from 89 to 53. If one repeats the triplet calculation, the Poisson
probability to observe two or more triplets is 0.003.

One of the lessons is the obvious one that one needs more data. The other
lesson is that one might raise the lower energy cut to enhance the
likelihood that the events come from a distance of less than 100 Mpc.
Reference to Figure 6 shows that 90$\%$ of the events with energy 
8 x 10$^{19}$ eV travel a distance of $\leq$ 100 Mpc. Thus, in searching
for pairs, a lower limit of 8 x 10$^{19}$eV might be a cut with some physical
motivation behind it. Such a selection is beyond the reach of the presently
operating experiments but may be within the reach of the Auger Observatory.
However Auger in the southern hemisphere looks away from the Virgo cluster
where there are significant concentrations of extragalactic matter which 
might serve as sources.

\subsection{Composition}

To measure the composition of the primary cosmic rays is the most difficult
challenge of all, far more so than the energies and directions. One seeks
to infer the nature of the primary particle from the 10$^{10}$ secondaries
produced. The two principal observables that can be traced to the
nature of the primary are the depth of maximum of the shower (X$_{max}$) and
the ratio of the muonic to to electromagnetic components of the shower.
There are secondary observables related to these primary observables.
For a given energy the showers are successively more penetrating (larger 
X$_{max}$) as one passes from a heavy primary to a proton to a photon.
A deeper shower has a sharper lateral distribution (the shower has less
distance to spread). The spread in time of shower particles that arrive at 
a detector far from the axis is larger for a deeply penetrating shower.
In addition, the muon to electromagnetic ratio decreases as the shower is
more penetrating. This ratio is roughly 40$\%$ lower for protons than for
the heaviest nucleus expected in the cosmic rays. Photons at the highest
energy $\geq$ 10$^{19}$ eV have a muon to electromagnetic ratio more than
a factor three less than protons.
 
In all the literature concerning composition one speaks of protons and iron
as if these are the only possibilities. This is because these two primaries
represent the extremes. There is barely the means to even separate iron
and protons, so that any mixture of protons and nuclei can be fit in this
two component model. A measurement of X$_{max}$ is a quantity most directly
related to composition. A measurement of this quantity as a function of
energy is Linsley's elongation rate. The bounds of the elongation rate
must be calculated by simulation, and these can vary by 10's of gm/cm$^2$
so the absolute position of X$_{max}$ as a function if primary is quite
uncertain. The slopes of the boundaries are less sensitive to the interaction
models. A steepening or a flattening of the elongation 
rate indicates a change in composition towards a lighter or heavier mix
of nuclei.

\begin{figure}
\centerline{\hbox{\psfig{figure=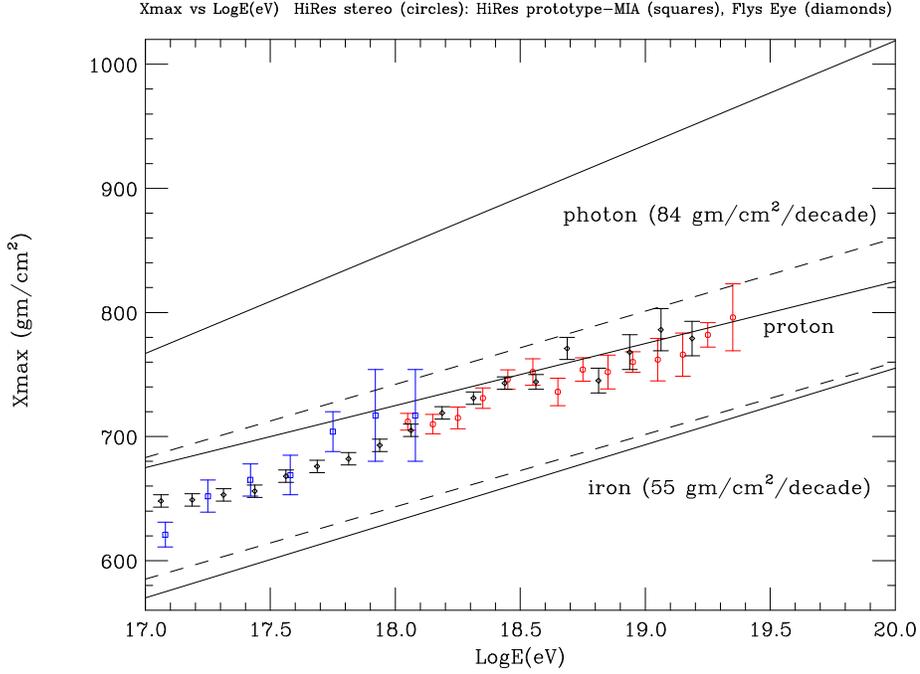,height=3.5in,angle=90}}}
\caption{Plot of X$_{max}$ vs energy measured by measurements of the Fly's Eye
and HiRes experiments. The boundaries indicated for iron and proton are
based on the QGSJET interaction model (solid) and the Sibyll model (dashed).
The elongation rate for photons is also plotted.}
\end{figure}

Additional composition information is contained in the fluctuation
of X$_{max}$. In the section on shower properties we saw that the 
fluctuation for
X$_{max}$ for protons was 53 gm/cm$^2$, while for iron it was 22 gm/cm$^2$.
The magnitude of these fluctuations is weakly dependent on the choice of
interaction model. 

The fluorescence detectors can measure X$_{max}$ with a statistical error of
$\leq$ 30 gm/cm$^2$. Recently the HiRes group presented a measurement of
X$_{max}$ in the range from 10$^{18}$ eV to 2 x 10$^{19}$ eV \cite {archbold}.
These results and prior measurements made with the HiRes prototype 
\cite{prototype}
and the original Fly's Eye experiment \cite{flyseye3} 
are plotted in Figure 36. Two different interaction models for the proton and
iron boundaries are indicated. While the boundary differences are significant,
it is amazing that the data do lie within the boundaries and the elongation 
rate for the different cases are about the same, 55 gm/cm$^2$ per decade.

\begin{figure}
\centerline{\hbox{\psfig{figure=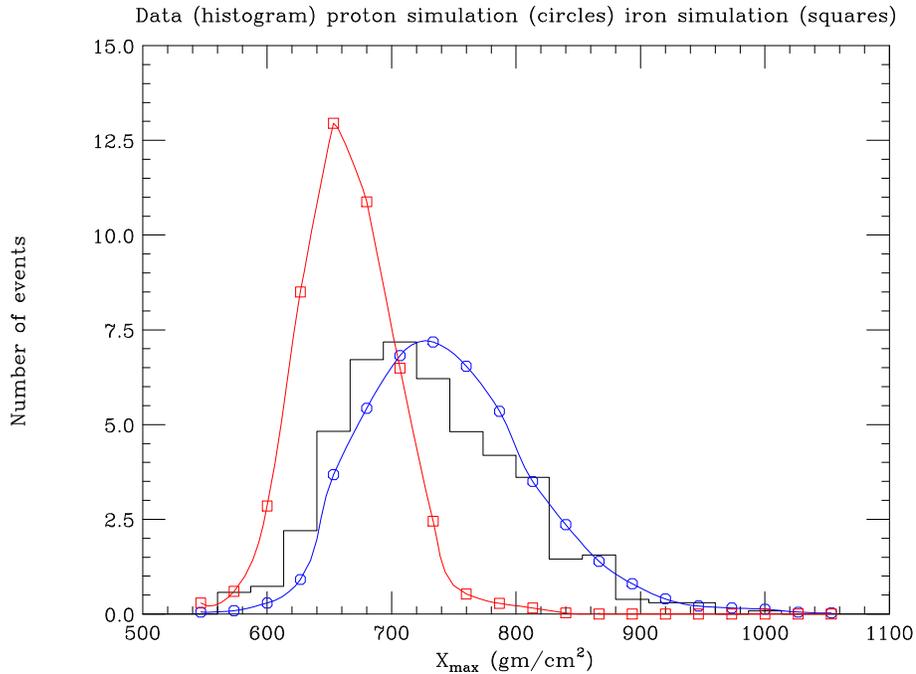,height=3.5in,angle=90}}}
\caption {Distribution of values of X$_{max}$ measured with the HiRes
stereo detector. The data are represented by the histogram. The expected
distribution for pure proton and pure iron are shown by the circles and
squares respectively.}
\end{figure}

In Figure 36 the elongation rate for photon showers is also plotted. Above
10$^{19}$ eV the curve strongly deviates from the indicated one because of
interaction in the Earth's geomagnetic field and the LPM effect. 
In either case the photon showers are much more deeply 
penetrating \cite{lpm}.

The interpretation of the data of Figure 36 is that the composition is
moving towards a lighter mixture. However the conclusion of an extremely 
light mixture
at 10$^{19}$ eV or a heavier mixture depends on the choice of the interaction
model. If
the boundary comes from the Sibyll model, the composition is less proton
rich than for the case of QGSJET.

Complementary information can be obtained from the fluctuations of X$_{max}$.
The fluctuation of  the HiRes stereo X$_{max}$ measurements 
for all of the energies lumped together is plotted in Figure 37. 
It would have made more sense
to divide the fluctuation distributions into separate energy bins so that
the evolution from a heavier composition to a lighter composition could have
been observed  through the broadening of the fluctuation distribution. With
the limited amount of data the HiRes group probably chose to combine all
the data and any situation can be simulated. 
The data are compared with the QGSJET simulation for pure iron and pure proton.
The shape of the X$_{max}$ fluctuation is very suggestive of a proton rich
composition.  The danger is that, if the distribution is broadened by errors
unaccounted for, there will be a bias towards a lighter composition. 
Nevertheless the relation between the width of the fluctuation distribution
and the nuclear species is almost independent of the interaction model, and 
it represents the most
promising measure of composition in the case where the errors  in the
X$_{max}$ measurement
have a negligible effect on the fluctuations.

\begin{figure}
\centerline{\hbox{\psfig{figure=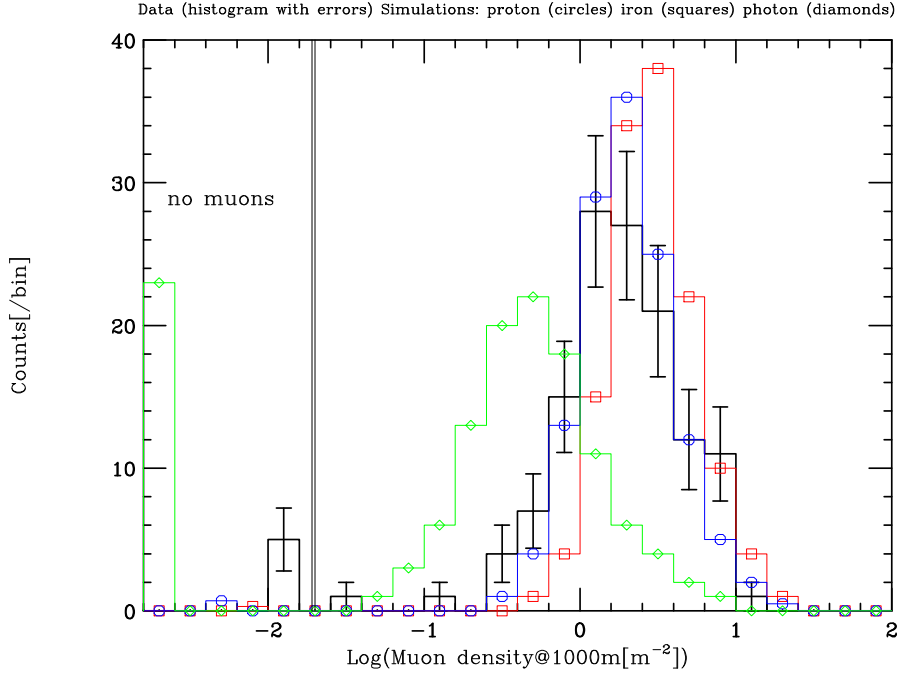,height=3.5in,angle=90}}}
\caption{Plot of muon density at 1000m for showers with energy 
$\geq$10$^{19}$ eV measured by the AGASA group. Black histogram with errors
is the data, histograms marked with squares, circles, and diamonds are
respectively the simulated distribution for iron, proton, and photon
primaries respectively.}
\end{figure}

An alternative method for the measurement of composition depends on the 
muon content of the showers. All simulations show that showers produced by
iron nuclei contain more muons than proton initiated showers. A typical
ratio is 1.4.
The AGASA array is outfitted with a number of muon
detectors in the southern part of their array. In a fraction of the events 
above 10$^{19}$ eV the density of muons at 1000m from the shower axis, 
$\rho_{\mu}$(1000), can be measured. The distribution for this quantity  
for iron, proton, and photon can be simulated.
These results are shown in Figure 38. The individual
measurements which each have a fractional RMS error of 40$\%$ scatter a 
great deal but their average is consistent with a proton rich mixture.
Although this conclusion agrees with the HiRes stereo measurement it is 
much more model dependent. There is no evidence for a photon component
and the upper limit for the photon fraction at 95$\%$ confidence is quoted
as 34$\%$.

Analysis of the past Haverah Park data using modern simulation techniques
has contributed significantly to
the question of the composition \cite{haverah1}. Using the steepness of
the lateral distribution, which is related to X$_{max}$, the Haverah Park 
group finds a proton
fraction of $\sim$ 40$\%$ in the range of energy 3 x 10$^{17}$ to 10$^{18}$
eV. This result is consistent with the view that the composition becomes
lighter as the energy increases from 10$^{17}$ to 10$^{19}$ eV. By comparing
the rates of vertical and horizontal showers the Haverah Park group has 
shown that the photon fraction at 10$^{19}$eV is less than 
40$\%$ \cite{haverah2}.   
 
There is an unsubstantiated feeling by many that at the extreme energies the
composition is proton rich. But at present the evidence is not very strong.
As more HiRes stereo data become available with a well controled
absolute energy calibration and with atmospheric attenuation corrections based
on nightly measurements and not averages, the fluorescence method for 
composition determination can become the most reliable. 
In addition if in passing from 10$^{17}$ eV to 10$^{19}$ eV the composition 
moves from heavy to light, one
must observe a narrow fluctuation distribution evolving 
into a broader one as theenergy increases.

\section{Conclusions}

I have attempted in this survey to explain without excessive complication the
important aspects for the measurement of the properties of cosmic rays
at the very highest energies ($\geq$ $10^{19}$ eV). There remain major
uncertainties in all areas, spectrum, anisotropy, and composition. The AGASA
experiment is now complete and one awaits the final catalog of events.
It should be possible for the AGASA group to extend their published 
measurements to larger zenith angles. I eagerly await the additional data
and improved analysis from the HiRes group.  

I am also looking forward to see the new data which will come from 
the Pierre Auger Observatory, now under construction in the southern hemisphere.
It is a hybrid detector which consists 
of both surface detectors and fluorescence telescopes. About 10$\%$ of showers 
will be observed simultaneously by both techniques. I will not list the 
advantages of the hybrid detector here; the reader can surely appreciate them.
The curious reader is encouraged to visit the many web sites devoted to
the Auger Observatory which can all be reached through {\bf www.auger.org}. 
As I write these conclusions (February  2004)
the Auger Observatory is operating with 220 surface detectors and 6 
30$^{\circ}$ x 30$^{\circ}$ fluorescence telescopes. By the end of 2004 there
will be 700-800 surface detectors and 12-14 fluorescence telescopes.
By the end of 2005 it should be complete with 1600 surface detectors 
covering 3000 km$^2$ and 24 fluorescence telescopes overlooking the
array.

The reader can surely appreciate the
enormous gain in statistics and the detailed information that will be 
obtained for
each event. However, from long experience in physics I succumb to caution
and will refrain from
making too many claims of what the Observatory will accomplish. Certainly
we will learn a great deal. Pierre Auger made his great discovery 66 years
ago. Nature rewarded John Linsley with a 10$^{20}$ eV shower 42 years ago.
I hope that the many questions that these great discoveries have raised
will be answered in a time much shorter than 42 or 66 years.

\section{Acknowledgements}

I am indebted to so many individuals over the years for contributing to my
understanding of the highest-energy cosmic rays. A partial list includes
Alan Watson, Michael Hillas, Tom Gaisser, M. Nagano, M. Teshima, 
Angela Olinto, Rene Ong, 
Johannes Knapp, Maria Teresa Dova,
Pierre Billoir, Tokonatsu Yamamoto, Maximo Ave, Paul Sommers, Brian Fick, 
G. Sigl, Murat Boratav, Paul Mantsch, and Simon Swordy. I want to thank 
A. Watson, P. Sommers, K. Arisaka, and M. Dova for a critical reading of the
manuscript but they are not responsible for any remaining errors.

This work was supported by the NSF grant PHY-0103717 and the Center for 
Cosmological Physics (NSF grant PHY-0114422).


\begin{thebibliography}{99}

\bibitem{nw00} M. Nagano, and A. A. Watson,
{\em Rev. Mod. Phys.}, {\bf 72}, (2000), 689. 

\bibitem{Linsley1}J. Linsley, {\em Phys. Rev. Lett.}, {\bf 10}, (1963), 146.  

\bibitem{Linsley2} J. Linsley, {\em 15th ICRC, Plovdiv, Bulgaria} 
 {\bf 12}, (1977), 89.

\bibitem{Hillas1} A. M. Hillas, {\em Ann. Rev. Astron. Astrophys.}, {\bf22},
(1984), 425.

\bibitem{Greisen} K. Greisen, {\em Phys. Rev. Lett.}, {\bf 16}, (1966), 748;
G. T. Zatsepin and V. A. Kuzmin, {\em Sov. Phys. JETP Lett. (Engl. Transl.)},
{\bf 4}, (1966), 78.

\bibitem{Olinto} D. DeMarco, P. Blasi, and A. V. Olinto, astro-ph/0301497,
to be published, {\em Astropart. Phys.}

\bibitem{Kronberg1} T. E. Clarke, P. P. Kronberg, and H. B\"{o}hringer,
astro-ph/0011281

\bibitem{Kronberg2} P. P. Kronberg, {\em Rep. Prog. Phys.}, {\bf 57}, (1994),
325; P. Blasi, S. Burles, and A. V. Olinto, {\em Ap. J.}, {\bf 514}, (1999),
L79.

\bibitem{Cronin} These calculations are not intended to replace serious
studies of propagation in magnetic fields given in the following references:
O. Deligny, A. Letessier-Selvon, and E. Parizot, astro-ph/0303624; G. Sigl,
M Lemoine, and P. Biermann, {\em Astropart. Phys.}, {\bf10}, (1999), 141; 
C. Isola, M Lemoine, and G. Sigl, {\em Phys. Rev. D}, {\bf 65}, (2002), 023004;
T. Stanev, R. Engel, A. Mucke, R. J. Protheroe, and J. P Rauchen, {\em Phys.
Rev. D}, {\bf 62}, (2000), 093005; C. Isola and G. Sigl, {\em Phys. Rev. D},
{\bf 66}, (2002), 083002; P. Blasi and A. V. Olinto, {\em Phys. Rev. D}, 
{\bf59}, (1999), 023001.

\bibitem{Simulations}The two simulation programs that are being used are
CORSIKA and AIRES. Information on CORSIKA can be found at 
{\bf www-ik3.fzk.de/$\sim$heck/corsika/}: information on AIRES can be
found at  {\bf www.fisica.unlp.edu.ar/auger/aires}.


\bibitem{Auger} A comprehensive description of the Pierre Auger Observatory
is given in {\em Proceedings of the 27th ICRC, Hamburg, Germany},
{\bf 2}, (2001), 699-784.


\bibitem{Ave} M. Ave for the Pierre Auger Collaboration, astro-ph/0308523.

\bibitem{Gaisser-Hillas} T. Gaisser and A. M. Hillas, {\em Proceedings of
the 15th ICRC, Plovdiv, Bulgaria}, {\bf 8}, (1977), 353



\bibitem{agasa1} K. Shinozaki et al., {\em Proc. 28th ICRC, Tokyo}, 
(2003), 401, Universal Academy Press, Inc.  

\bibitem{watson} The Haverah Park array consisted of deep water tanks;
see for example, M. A. Lawrence, R. J. O. Reid, and A. A. Watson,
{\em J. Phys.} {\bf G17}, (1991), 733.


\bibitem{Linsley3} Linsley was the first to observe the spread of arrival times
as a function of distance from the shower axis: see J. Linsley and L. Scarsi,
{\em Phys. Rev.}, {\bf 128}, (1962), 2384. Experiments performed before the
advent of  microelectronics recorded data with oscillocope photography and
 risetime information was available for the analysis. See for example,
R. Walker and A. A. Watson, {\em J. Phys G}, {\bf 8}, (1982), 1131.

\bibitem{agasa2} M. Takeda et al., {\em Astro. Part. Phys.}, {\bf 19}, (2003),
499.

\bibitem{agasa3} M. Takeda et al., {\em Proc. 28th ICRC, Tokyo}, (2003), 381,
Universal Academy Press, Inc.


\bibitem{Hillas2} A. M. Hillas, D. J. Marsden, J. D. Hollows, and H. W. Hunter,
{\em Proc. 12th ICRC, Hobart}, {\bf 3}, (1971), 1001, University of Tasmania

\bibitem{flyseye1} D. Bird et al., {\em Astrophys. J.}, {\bf 511}, (1999), 739.

\bibitem{Bunner} A. N. Bunner, Ph.D. thesis, Cornell University (1964);
a copy of his plot is reproduced in R. M. Baltrusaitis et al., {\em Nucl.
Instrum. Methods Phys. Res. A}, {\bf 240}, (1985), 410. The most recent
measurements of the total fluorescence yield are by, F. Kakimoto et al.,
{\em Nucl. Instrum. Methods Phys. Res. A},{\bf 372}, (1996), 527, and
M. Nagano, K. Kobayakawa, N. Sakaki, and K. Ando, {\em Astropart. Phys.},
{\bf 20}, (2003), 293.
 

\bibitem{matthews1} J. A. J. Matthews, { \em Proc. SPIE Conference on
Astronomical Telescopes and Instrumention}, (2002), Waikoloa, Hawaii; 
this paper describes the calibration of the Auger fluorescence telescope.
The problems and techniques are common to both the Auger and HiRes detectors.
A copy of this paper is also available as GAP-2002-029 from Technical 
Information at {\bf www.auger.org}.

\bibitem{flyseye2} D. Bird et al., {\em Nucl. Instrum. Methods Phys. Res. A},
{\bf 349}, (1994), 592.

\bibitem{sommers1} P. Sommers, {\em Astropart. Phys.}, {\bf 3}, (1995), 349.


\bibitem{matthews2} J. A. J. Matthews and R. Clay, {\em Proc. 27th ICRC,
Hamburg, Germany}, {\bf 2}, (2001), 745; also available as GAP-2001-029
from Technical Information at {\bf www.auger.org}.

\bibitem{abu-zayyad} R. U. Abassi et al. for the The High Resolution Flys 
Eye Collaboration, astro-ph/0208243 v7 (2003); a brief version of this 
paper was presented at 
the 2003 ICRC; D. Bergman et al., {\em Proc. 28th ICRC, Tokyo}, (2003) 397, 
Universal Academy Press, Inc. 


\bibitem{springer} W. Springer et al., {\em Proc. 28th ICRC, Tokyo}, (2003),
409, Universal Academy Press, Inc; the HiRes stereo spectrum presented here
is not contained in this paper but was presented at the conference.

\bibitem{bahcall} J. N. Bahcall and E. Waxman, {\em Phys. Lett.}, {\bf B556},
(2003), 1.

\bibitem{agasa4} M. Teshima et al., {\em Proc. 28th ICRC, Tokyo}, (2003),
437, Universal Academy Press, Inc. The catalog of the 57 events is given by
N. Hayashida et al., astro-ph/0008102, (2000).


\bibitem{haverah} The Haverah Park directions were provided by A. A. Watson,
private communication. Recently the Haverah Park energy scale was reduced by
about 30$\%$ and that change has been roughly accounted for in the selection of
events for Figure 33. see M. Ave et al. {\em Proc 27th ICRC, Hamburg} {\bf 1},
(2001), 381.

\bibitem{yakutsk} The directions were provided by A. A. Mikhailov. I wish
to thank Professor A. V. Glushkov for permission to use the Yakutsk directions
and energies. The selection of Yakutsk events for Figure 33  were reduced 
by a factor log$_{10}$=0.1 to better match the AGASA spectrum.

 
\bibitem{volcano} J. Linsley, {\em Catalog of Highest Energy Cosmic Rays},
edited by M. Wada (World Data Center of Cosmic Rays, Institute of Physical
and Chemical Research, Itabashi, Tokyo) {\bf 1}, (1980), 1. 

\bibitem{feldman} The 68$\%$ limits were taken from the tables of
R. D. Cousins and G. J. Feldman, {\em Phys. Rev. D} , { \bf 57}, (1998), 3873.

\bibitem{archbold} G. Archbold and P. V. Sokolsky, for the HiRes collaboration.
{\em Proc. 28th ICRC, Tokyo}, (2003), 405, Universal Academy Press, Inc.

\bibitem{prototype} T. Abu-Zayyad et al., {\em Ap. J.}, {\bf 557}, (2001), 686.

\bibitem{flyseye3} D. J. Bird et al., {\em Phys. Rev. Lett.}, {\bf 71}, (1993),
3401. The experimental points from the original Fly's Eye experiment were
raised by 20 gm/cm$^2$ to compensate for a systematic experimental bias.
This is described in T. K.Gaisser et al., {\em Phys. Rev. D}, {\bf47}, (1993),
1919. We quote from page 1925 of that paper: `` We estimate that the simulated
showers can be shifted to a shallower X$_{max}$ to a maximum of about 20
gm/cm$^2$, or alternatively the experimentally detected showers can be assigned
X$_{max}$ deeper by the same amount.''

\bibitem{lpm} There are many papers on the interaction of photons with
energy $\geq$ 10$^{19}$ eV with the geomagnetic field and the LPM effect.
 A selection of these: T. Stanev and H. P. Vankov,
{\em Phys. Rev. D} {\bf 55}, (1997), 1365; X. Bertou, P. Billoir, and S.
Dagoret-Campagne, {\em Astropart. Phys.}, {\bf 14}, (2000), 121; H. P. 
Vankov, N. Inoue, and K. Shinozaki, {\em Phys. Rev. D}, {\bf 67}, (2003),
043002; P. Homola et al., astro-ph/0311442.

\bibitem{haverah1} M. Ave et al., {\em Astropart. Phys.}, {\bf 19}, (2003),
61.

\bibitem{haverah2} M. Ave et al., {\em Phys. Rev. Lett.}, {\bf 85}, (2000),
2244.


\end{thebibliography}
\end{document}